\documentclass[aps,reprint, prb, longbibliography, superscriptaddress, showpacs,nobibnotes,floatfix]{revtex4-1}
\usepackage{graphicx}
\usepackage{bm,amsmath}
\usepackage{dcolumn}
\usepackage{natbib,xcolor}
\usepackage[version=4]{mhchem}
\usepackage{upgreek}

\usepackage[normalem]{ulem}

\begin{document}
\title{Electronic structure of the Kramers nodal-line semimetal YAuGe}

\author{Takashi Kurumaji}
\affiliation{Division of Physics, Mathematics and Astronomy, California Institute of Technology, Pasadena, California 91125, USA}
\affiliation{Department of Advanced Materials Science, University of Tokyo, Kashiwa 277-8561, Japan}
\author{Jorge I. Facio}
\affiliation{Centro Atomico Bariloche, Instituto de Nanociencia y Nanotecnologia (CNEA-CONICET) and Instituto Balseiro, Av. Bustillo 9500, Argentina}
\author{Natsuki Mitsuishi}
\affiliation{RIKEN Center for Emergent Matter Science (CEMS), Wako 351-0198, Japan}
\author{Shusaku Imajo}
\affiliation{Institute for Solid State Physics, University of Tokyo, Kashiwa, Chiba 277-8581, Japan}
\author{Masaki Gen}
\affiliation{Institute for Solid State Physics, University of Tokyo, Kashiwa, Chiba 277-8581, Japan}
\author{Motoi Kimata}
\altaffiliation[Present affiliation: ]{Advanced Science Research Center, Japan Atomic Energy Agency, Tokai, Ibaraki 319-1195, Japan}
\affiliation{Institute for Materials Research, Tohoku University, Sendai, Miyagi, 980-8577, Japan}
\author{Linda Ye}
\affiliation{Division of Physics, Mathematics and Astronomy, California Institute of Technology, Pasadena, California 91125, USA}
\author{David Graf}
\affiliation{National High Magnetic Field Lab, Tallahassee, Florida, 32310 USA}
\author{Masato Sakano}
\affiliation{Quantum-Phase Electronics Center and Department of Applied Physics, University of Tokyo, Bunkyo-ku, Tokyo 113-8656, Japan}
\author{Miho Kitamura}
\affiliation{Photon Factory, Institute of Materials Structure Science, High Energy Accelerator Research Organization (KEK), Tsukuba 305–0801, Japan}
\author{Kohei Yamagami}
\affiliation{Japan Synchrotron Radiation Research Institute, Sayo 679–5198, Japan}
\author{Kyoko Ishizaka}
\affiliation{RIKEN Center for Emergent Matter Science (CEMS), Wako 351-0198, Japan}
\affiliation{Quantum-Phase Electronics Center and Department of Applied Physics, University of Tokyo, Bunkyo-ku, Tokyo 113-8656, Japan}
\author{Koichi Kindo}
\affiliation{Institute for Solid State Physics, University of Tokyo, Kashiwa, Chiba 277-8581, Japan}
\author{Taka-hisa Arima}
\affiliation{Department of Advanced Materials Science, University of Tokyo, Kashiwa 277-8561, Japan}
\affiliation{RIKEN Center for Emergent Matter Science (CEMS), Wako 351-0198, Japan}
\date{\today}
\begin{abstract}
Nodal-line semimetals are a class of topological materials hosting one dimensional lines of band degeneracy.
Kramers nodal-line (KNL) metals/semimetals have recently been theoretically recognized as a class of topological states inherent to all non-centrosymmetric achiral crystal lattices.
We investigate the electronic structure of candidate KNL semimetal YAuGe by angle-resolved photoemission spectroscopy (ARPES) and quantum oscillations as well as by density functional theory (DFT) calculations.
DFT has revealed that YAuGe hosts KNLs on the $\Gamma$-A-L-M plane of the Brillouin zone, that are protected by the time reversal and mirror-inversion symmetries.
Through ARPES and quantum oscillations we identify signatures of hole bands enclosing the $\Gamma$ point, and the observed splitting of quantum oscillation frequency with angle is attributed to spin-orbit-coupling-induced band splitting away from the KNLs.
Furthermore, we show that the degeneracy of the nodal lines along the $\Gamma$-A line is lifted by the time-reversal-symmetry breaking when the Y is substituted by magnetic $R$ ions ($R$ = rare earth).
This becomes a source of Berry curvature and contributes to the anomalous Hall effect in magnetic $R$AuGe.
These findings establish $R$AuGe as a new class of KNL semimetals offering significant potential for engineering of anomalous magnetotransport properties via magnetic rare-earth substitution.
\end{abstract}

\keywords{Kramers nodal-line semimeal, Quantum oscillations, Dirac point}
\maketitle

\section{Introduction}
Nodal-line semimetals, where electron band crossings form one-dimensional lines in the momentum space, offer a platform for exploration of a variety of novel phenomena, such as drumhead surface states as well as of enhanced quantum transport and optical responses ~\cite{burkov2011topological,liu2018giant,fang2016topological}.
Nodal lines have played a key role to understand the existence of Weyl points in a number of materials.
In this context, nodal lines characterize the touching of valence and conduction bands in the absence of spin-orbit interaction (SOI) and become split in the relativistic case
\cite{fang2016topological}.
More recently, Kramers nodal-line (KNL) semimetals were introduced in Ref. \onlinecite{xie2021kramers}, where it was shown that all time-reversal symmetric noncentrosymmetric achiral materials are expected to host well-defined nodal lines, even in the presence of strong SOI.
These KNLs are protected by time-reversal symmetry together with an achiral symmetry, such as a reflection symmetry.
As a result, the degenerate states form lines that connect time-reversal invariant momenta (TRIMs) of the Brillouin zone (BZ).
A number of materials such as $T$RuSi ($T=$ Ti, Nb, Hf, Ta) ~\cite{shang2022unconventional}, $R$Te$_3$ ($R=$ La, Y) ~\cite{sarkar2023charge,sarkar2024kramers}, and SmAlSi ~\cite{zhang2023kramers} are reported to host KNL from photoemission spectroscopy, while clear transport signatures associated with the KNL have not been well established yet.

\begin{figure*}[t]
	\includegraphics[width =  \textwidth]{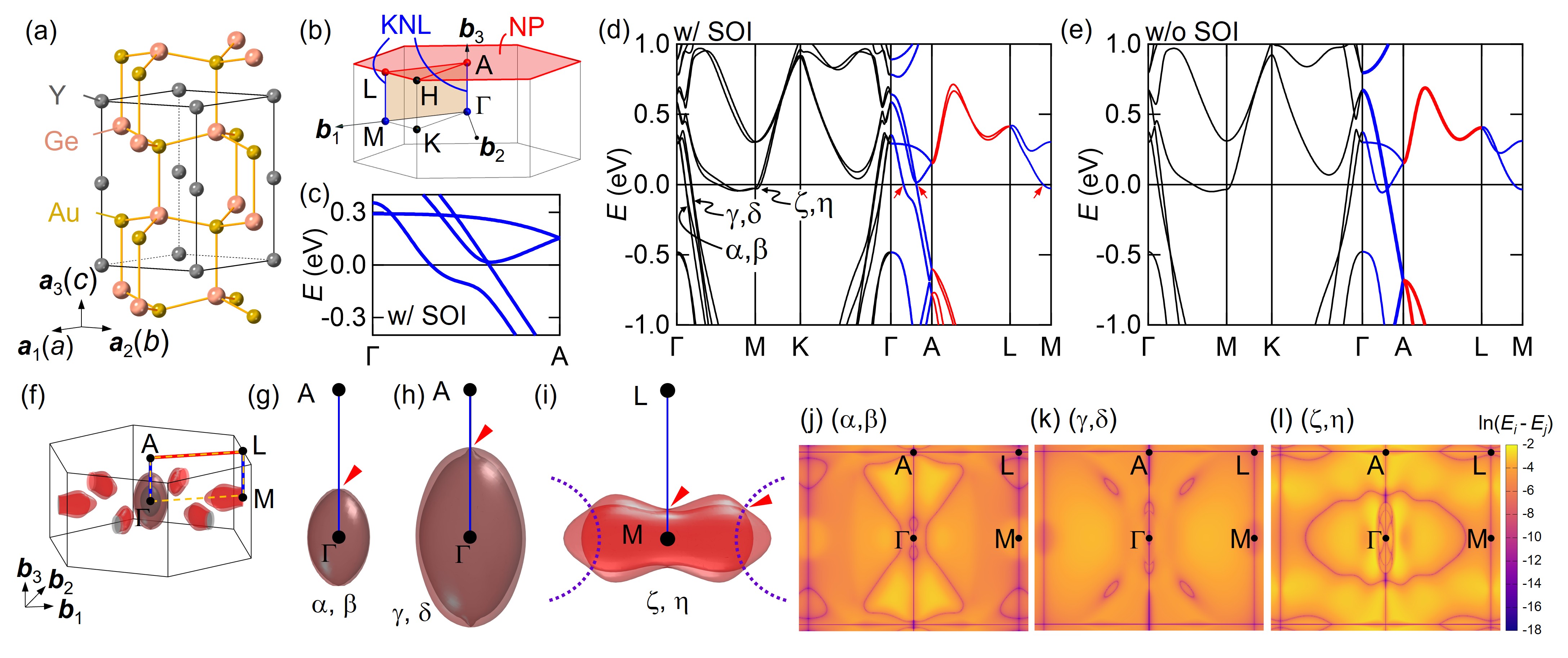}
	\caption{\label{figBand}(a) Crystal structure of YAuGe. 
 (b) BZ and high symmetry points.
 $\Gamma$, A, M, and L are TRIMs.
 Blue lines represent the KNLs, red hexagonal plane ($k_z=\pi/c$) is the NP, and orange area is invariant under the mirror inversion ($m_{010}$).
(c) Band dispersion along the $\Gamma$-A line in YAuGe, calculated with the SOI.
 All the bands are doubly-degenerate KNLs.
 (d)-(e) DFT calculations of the band structure in YAuGe (d) with SOI and (e) without SOI.
 In (d), bands along $\Gamma$-A and M-L lines form KNLs as denoted by blue curves, and red curves along A-L line are two-fold degenerate due to the NP nature.
 $\alpha$, $\beta$, $\gamma$, and $\delta$ provide hole pockets at $\Gamma$, and $\zeta$ and $\eta$ are electron pockets at M.
 Red arrows denote the position of the pinch point in FSs.
 In (e), thick curves along $\Gamma$-A and A-L lines denote four-fold degenerate bands.
 (f) Fermi surfaces of YAuGe obtained by DFT calculations.
 Gray (red) sheets are for the hole (electron) pockets.
 (g) ((h), (i)) Side view of the Fermi surfaces for $\alpha$ and $\beta$ ($\gamma$ and $\delta$, $\zeta$ and $\eta$) branches.
 Blue line (red triangle) denotes KNL (pinch point).
 For (i), accidental NLs are also denoted by dashed purple curves, which gives additional four pinch points between $\zeta$ and $\eta$.
 (j)-(l) Color map of the band energy difference plotted in a log scale, i.e., $\ln{(E_i-E_j)}$ between (j) $i=\alpha$, $j=\beta$, (k)  $i=\gamma$, $j=\delta$, and (l) $i=\zeta$, $j=\eta$.
 Purple curves correspond to the NLs.
 }
\end{figure*}

$R$AuGe ($R=$ Sc, La-Nd, Sm, Gd-Tm, and Lu) is a family of the $RTX$ intermetallic phases ($T$ is a transition metal element while $X$ a $p$-group element) belonging to the polar achiral structure (space group: $P6_{3}mc$) as shown in Fig. \ref{figBand}(a) ~\cite{rossi1992ternary,pottgen1996crystal}, satisfying the conditions to host KNLs.
Au and Ge atoms form a wurtzite-type network which breaks inversion symmetry, and $R$ atoms form a triangular lattice in the $ab$ plane. 
The interplay between the polar nature and high electrical conductivity of epitaxially grown thin films has recently attracted attention ~\cite{du2019high,du2022controlling,laduca2024cold}, and it has been reported that the introduction of magnetic ions in the $R$ sites induces a large anomalous Hall effect ~\cite{ram2023multiple,kurumaji2024metamagnetism}.
The connection between these transport properties and the geometrical/topological characteristics of the electronic structure has not yet been studied to date.

Here we investigate the electronic structure of polar YAuGe via a combination of $\it{ab}$ $\it{initio}$ density functional theory (DFT) calculations, angule-resolved photoemission spectroscopy (ARPES), quantum oscillations in resistivity and magnetization torque, and specific heat.
DFT calculations reveal that the electron bands near Fermi energy ($E_{\text{F}}$) are of KNL nature along the $\Gamma$-A line.
In ARPES we identify hole pockets centered at the $\Gamma$ point, consistent with the DFT calculations and the dominant hole  carrier transport observed in Ref. \onlinecite{kurumaji2024metamagnetism}.
The highly mobile hole pockets exhibit quantum oscillations at high magnetic fields, and an observed splitting with angle is attributed to due to the underlying spin-orbit coupling in the system.
Substitution of magnetic $R$ atoms into Y site induces time-reversal symmetry breaking, which lifts the degeneracy of KNL along the $\Gamma$-A line.
This effect gives rise to the finite Berry curvature among adjacent bands associated with the KNL.
By comparing the experiments in $R$AuGe ($R=$ Dy, Ho, Er, Tm) with DFT calculations, we identified that both extrinsic and intrinsic mechanisms contribute to the observed AHE.
These findings demonstrate the presence of KNLs near the Fermi level in noncentrosymmetric $R$AuGe, and highlight their interplay with magnetic orders and time reversal symmetry.

\section{Results}
\subsection{Density functional theory calculations of electronic band structure of YAuGe}

We first discuss the KNL expected in YAuGe from the perspective of symmetry.
Figure \ref{figBand}(b) represents the first BZ of YAuGe.
The TRIMs are located at $\Gamma$, A, L, M, which are expected to be connected by the KNLs ~\cite{xie2021kramers}.
Due to the mirror-inversion symmetry, the KNLs are predicted to be confined on the $\Gamma$-A-L-M plane (orange plane in Fig. \ref{figBand}(b)) ~\cite{xie2021kramers}.
We note that the $6_3$ screw along the $c$ axis leads to an additional degeneracy (nodal plane, NP) on the BZ boundary plane at $k_z=\pi/c$ (red plane in Fig. \ref{figBand}(b)) ~\cite{chang2018topological}.
Symmetry arguments based on the representation theory are given in Appendix B.

The above symmetry consideration is well consistent with the calculated band structure of YAuGe.
Figure \ref{figBand}(c) shows the band structure with SOI along the $\Gamma$-A line near $E_{\text{F}}$.
We note that all the shown bands are doubly degenerate and that the degeneracy cannot be lifted by SOI, as expected for KNL.
Figures \ref{figBand}(d)-(e) compare the band structures with and without the SOI along high symmetry lines in the BZ highlighted in Fig. \ref{figBand}(b).
Along $\Gamma$-A, L-M, and A-L lines, the four-fold degeneracy is lifted by the introduction of the SOI, while the two-fold spin degeneracy is preserved, consistent with the symmetry considerations (see Appendix B).
In contrast, the two-fold spin degeneracy is fully lifted along $\Gamma$-M, and $\Gamma$-K lines.
Without SOI, near $\Gamma$ one can find two branches of doubly-degenerate hole bands (Fig. \ref{figBand}(e)) near $E_{\text{F}}$; upon the introduction of SOI (Fig. \ref{figBand}(d)), the noncentrosymmetric nature of the lattice splits each hole branch into two to form pairs of bands we assign as ($\alpha$, $\beta$), and ($\gamma$, $\delta$), respectively, while the degeneracy at $\Gamma$ remains as the KNL runs through $\Gamma$.
The same occurs around the M point, where the electron bands split into the ($\zeta$, $\eta$) pair.

The Fermi surfaces of KNL semimetals are predicted to host the so-called pinch points when a KNL crosses the Fermi level ~\cite{xie2021kramers}.
At such points, two spin-orbit-split FSs touch with each other as denoted by red arrows in Fig. \ref{figBand}(d).
Figures \ref{figBand}(f)-(i) depict the FSs in YAuGe.
The hole-type FSs for the ($\alpha$, $\beta$) and ($\gamma$, $\delta$) pairs (Figs. \ref{figBand}(g)-(h)) have two pinch points along $\Gamma$-A (see red triangle).
The electron pocket pair of ($\zeta$, $\eta$) as shown in Fig. \ref{figBand}(i) also has pinch points in the M-L line connecting TRIMs, while one can identify four additional pinch points, as a result of accidental degeneracies near the Fermi level represented by purple dashed lines in Fig. \ref{figBand}(i).

To clarify the trajectory of KNLs in the BZ, the energy difference of SOI-pair bands is plotted in the $\Gamma$-A-L-M plane (Figs. \ref{figBand}(j)-(l)).
As predicted by the symmetry, the KNLs are observed as straight lines connecting along $\Gamma$-A, and M-L lines.
We identify other branches of nodal lines forming self-connected loops on the mirror plane, which are related to the additional pinch points of the ($\zeta$, $\eta$) FSs.
Such accidental NLs are allowed when the bands belong to different IRs, and the NLs are confined at the $\Gamma$-M-L-A plane due to the same mechanism for the KNLs \cite{xie2021kramers}.
Although this is an interesting aspect of the electronic structure hosting KNLs, in principle, these loops are not classified as the KNL because they do not connect the TRIMs.

\subsection{Angle-resolved photoemission spectroscopy and quantum oscillation measurements of YAuGe}
\begin{figure}[t]
	\includegraphics[width =  \columnwidth]{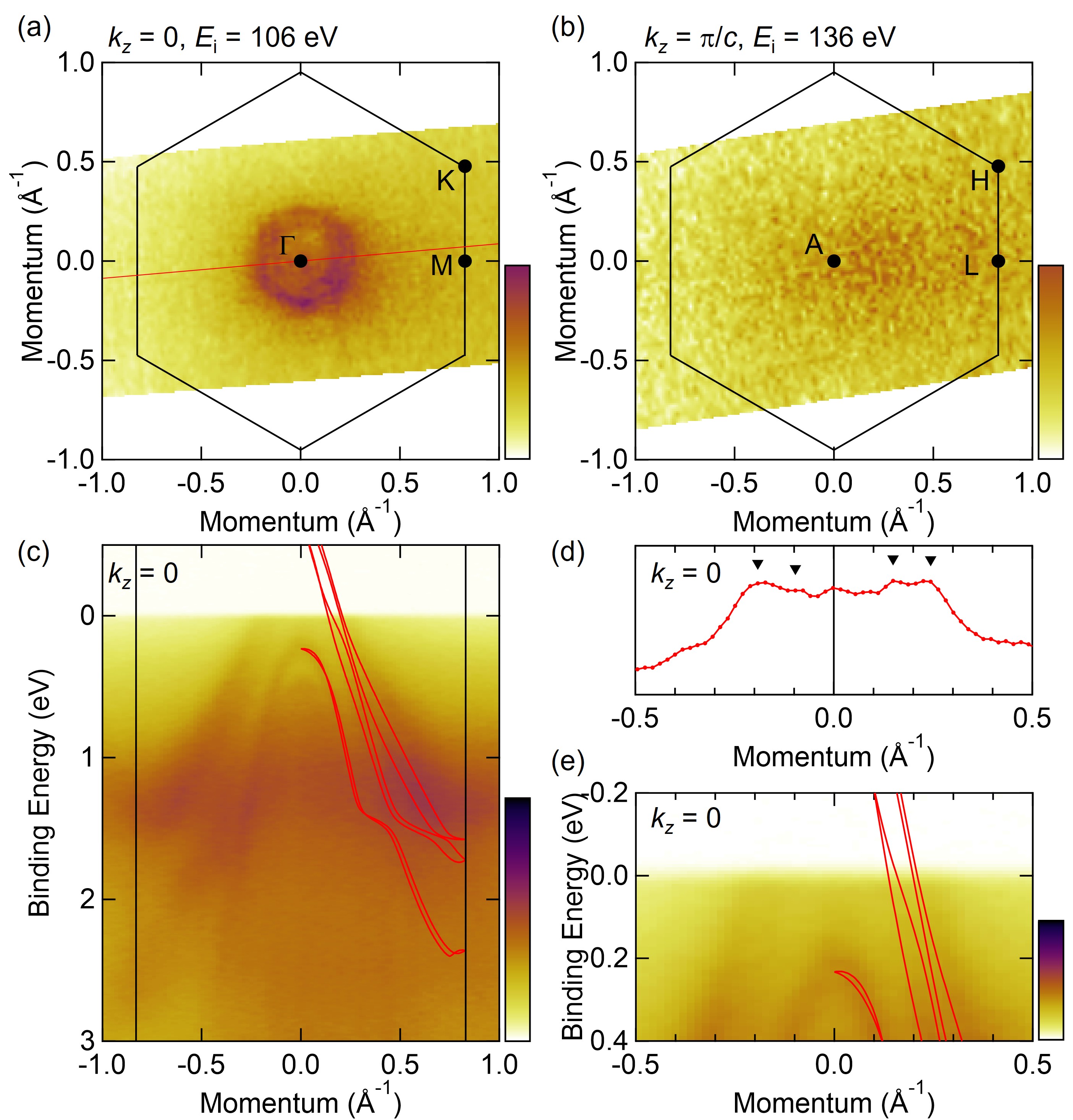}
	\caption{\label{figARPES} (a) [(b)] ARPES intensity plots at $E_{\text{F}}$ of YAuGe measured at 15 K with $E_{\text{i}}=106$ eV [136 eV] incident photons, which approximately probes the $\Gamma$-M-K [A-L-H] plane.
 The (001) surface BZ is marked with the black solid hexagon.
 (c) Band dispersion along the momentum cut marked by the red line in (a) (near M-$\Gamma$-M).
 (d) Intensity profile at $E_{\text{F}}$ extracted from the ARPES image (c).
 Black triangles mark the intensity peaks for the hole bands.
 (e) Zoom-in image of the band dispersion at $E_{\text{F}}$ for (c).
}
\end{figure}

Having confirmed via DFT calculations the presence of KNLs near $E_{\text{F}}$ in YAuGe, we investigate the electronic structure by ARPES.
Figures \ref{figARPES}(a)-(b) show the FS cross section collected at $T=15$ K using incident photons of $E_{\text{i}}=106$ eV and 136 eV, respectively, probing around the $k_z=0$ and $k_z=\pi/c$ planes.
We find isotropic Fermi pockets near the $\Gamma$ point in the $k_z=0$ plane.
The details of photon-energy-dependent ARPES results are provided in Appendix C.
Figure \ref{figARPES}(c) shows the band dispersion near the $\Gamma$-M lines as depicted by the red line cut in Fig. \ref{figARPES}(a).
We have identified hole pockets from the spectra, which agrees well with the energy dispersion calculated by the DFT calculations (red curves).
Here we shifted $E_{\text{F}}$ of the DFT calculation by -0.25 eV to overlap the band dispersions with the ARPES image.
Considering that the Fermi momenta obtained by DFT calculation agree well with SdH measurement (as discussed below and summarized in Table \ref{table}), this ARPES result might be suggesting the possible electron depletion occurring near the sub-surface region of the cleaved crystal.
Focusing on the bands near $E_{\text{F}}$ (Figs. \ref{figARPES}(d)-(e)), we can identify signatures of two branches of bands that can be attributed to the two groups ($\alpha$, $\beta$) and ($\gamma$, $\delta$), while the SOI splitting within each pair appears to fall below experimental resolution.

\begin{figure}[t]
	\includegraphics[width =  \columnwidth]{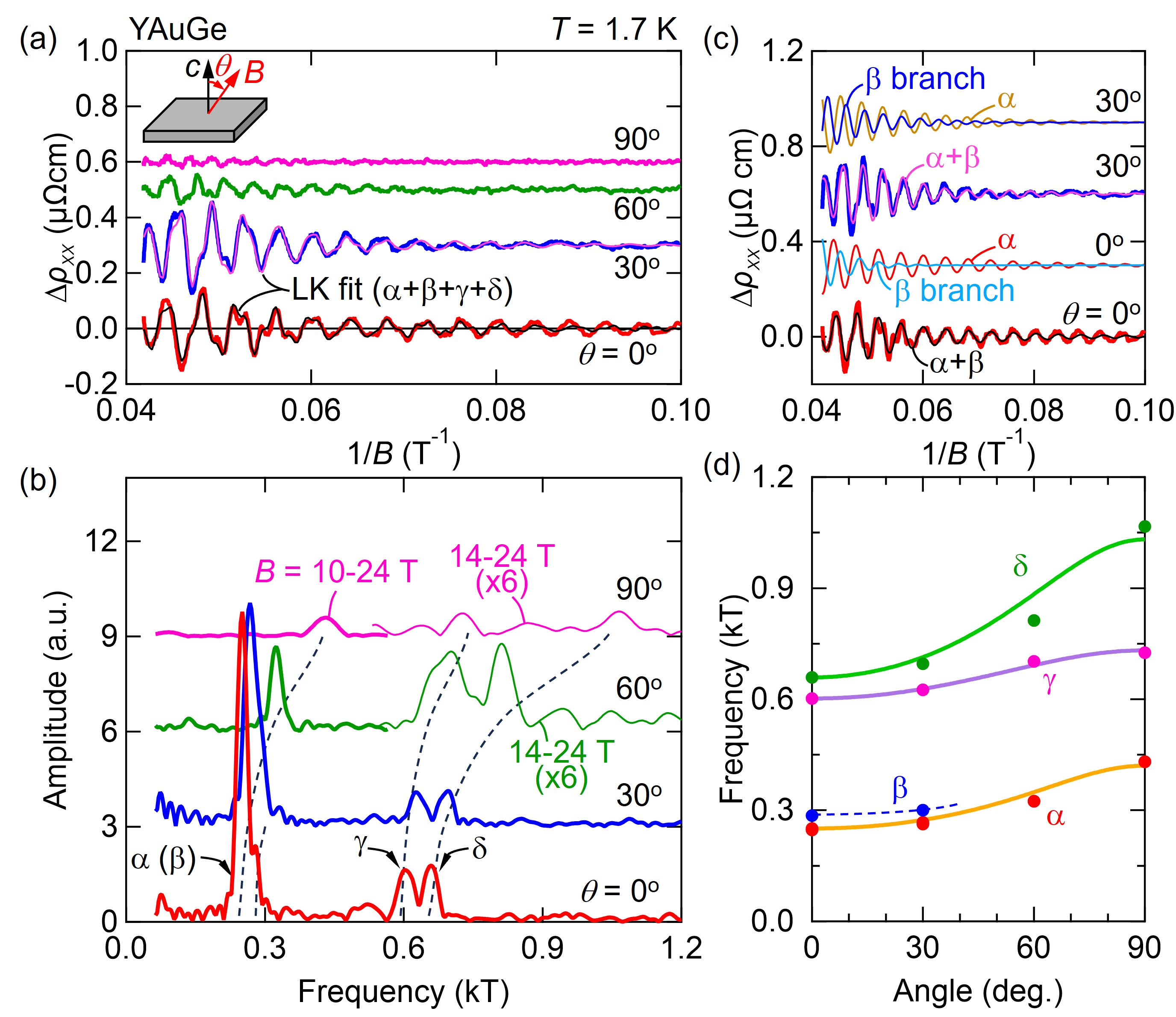}
	\caption{\label{figAng} (a) Inverse-field ($1/B$) dependence of the background-subtracted resistivity ($\Delta \rho_{xx}$) measured at $T =$ 1.7 K with the $B$ tilted from the $c$ axis to the $ab$ plane by $\theta$ (see inset).
 Thin black (pink) curve is the LK fit using Eq. (\ref{LKfull}), where four branches (from $\alpha$ to $\delta$) are considered.
 (b) FFT of SdH oscillations at $T = 1.7$ K with various $\theta$ analyzed with different $B$ windows.
 Thick: $B = 10$ to 24 T; thin: $B = 14$ to 24 T.
 Positions of the branches $\alpha$, $\gamma$, and $\delta$ are assigned.
 The $\beta$ branch is not resolved from the peak for $\alpha$.
 Dashed curve is the guide to eyes.
 (c) Comparison between SdH oscillations at $\theta=0,30^{\circ}$ and the LK fit.
 For $\theta=0^{\circ}$, thin black curve is the summation of the oscillation components for $\alpha$ and $\beta$ and thin red (cyan) curve above is the individual oscillation components for $\alpha$ ($\beta$).
 For $\theta=30^{\circ}$, thin pink curve is the summation of the oscillation components for $\alpha$ and $\beta$ and thin yellow (blue) curve above is the individual oscillation components for $\alpha$ ($\beta$).
 (d) Angular ($\theta$) dependence of the SdH oscillation frequency for each branch (closed circles).
 Solid (dashed) lines are fits with the model of an elliptical Fermi surface (guide to eyes).
}
\end{figure}

To better resolve the SOI splittings in the Fermi surfaces, we turn to Shubnikov-de Haas oscillations in the resistivity of YAuGe at a high magnetic field up to 24 T.
Figure \ref{figAng}(a) shows quantum oscillations in $\Delta \rho_{xx}$ after subtraction of a smooth background. 
Fast Fourier transformation (FFT) of $\Delta \rho_{xx}$ reveals the oscillation frequencies as peaks (Fig. \ref{figAng}(b)).
At $\theta=0$, compared with the DFT calculations, we identify three branches $\alpha$ ($F_{\alpha}=$ 250 T), $\gamma$ ($F_{\gamma}=$ 602 T), and $\delta$ ($F_{\delta}=$ 659 T), while the expected peak for $\beta$ overlaps with that of  $\alpha$.
The temperature dependence of oscillation amplitude is analyzed with the Lifshitz-Kosevich (LK) formula (see Appendix D), which gives the effective mass: $m^*_{\alpha}=0.10m_0$, $m^*_{\gamma}=0.21m_0$, $m^*_{\delta}=0.25m_0$, where $m_0$ is the free electron mass.
The consistency with the DFT calculations is also confirmed (see Fig. \ref{figSBand}).

As magnetic field is tilted from the $c$ axis towards the $ab$ plane, all the oscillation peaks shift to higher frequencies (Fig. \ref{figAng}(b)), suggesting that the FSs are all elliptically elongated along the $k_z$ direction.
We note that the oscillations associated with the electron pockets around M ($\zeta$ and $\eta$) are not observed at any field angles possibly due to the large effective mass, which can also be inferred from the band dispersion in Fig. \ref{figBand}(d) (see also Fig. \ref{figSBand}(b)).
From specific heat (see Appendix D), we estimate the effective mass of electron pockets to be $m^*_{\zeta,\eta}\sim 0.4m_0$.
We note that this is the averaged value of the effective mass tensor anisotropic in the momentum space, and the heavier mass than that for hole bands is consistent with DFT.
We also note that in the previous study by some of the authors ~\cite{kurumaji2024metamagnetism} the electron carrier mobility was estimated to be around $\mu_e =100$ cm$^2$/Vs by a two-carrier analysis, which requires $B\sim 100$ T to resolve quantum oscillations ($\mu_eB\sim 1$).

\begin{figure*}[t]
	\includegraphics[width =  0.7\textwidth]{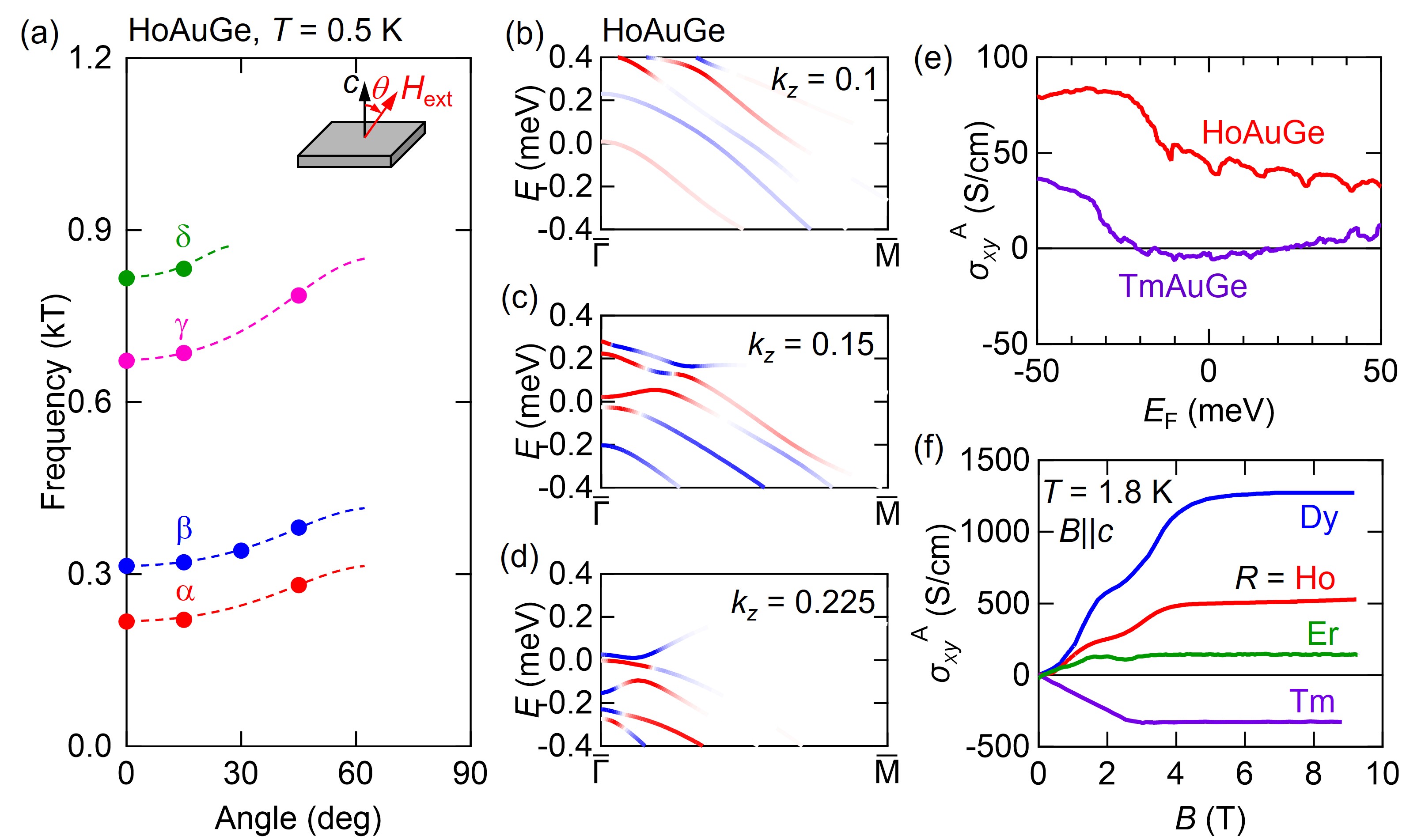}
	\caption{\label{figBerry} (a) Angular ($\theta$) dependence of the SdH oscillation frequency for each branch (closed circles) observed at $T=0.5$ K in HoAuGe.
 Dashed lines are guide to eyes.
 (b)-(d) Berry curvature in HoAuGe at different $k_z$.
 (e) AHC ($\sigma^{\text{A}}_{xy}$) for HoAuGe and TmAuGe calculated with open core (see methods).
 (f) Magnetic field dependence of AHC observed in $R$AuGe ($R=$ Dy-Tm) for $B\parallel c$ at $T=1.8$ K.
 For $R=$ Dy and Ho, $\sigma^{\text{A}}_{xy}$ is reproduced from Ref.~\onlinecite{kurumaji2024metamagnetism}.
 }
\end{figure*}

Although not clearly resolved in the FFT spectra shown in Fig. 3(b), the coexistence of the $\alpha$ and $\beta$ branches is indicated by the nonmonotonic evolution of oscillation amplitudes as a function of inverse field (see $\theta = 0^{\circ}$ and $\theta = 30^{\circ}$ in Fig. \ref{figAng}(a)), suggesting a beating pattern between the closely spaced $\alpha$ and $\beta$ branches.
In order to resolve these two branches, we fit the $\Delta \rho_{xx}$ vs $1/B$ with the LK formula as below ~\cite{shoenberg2009magnetic}
\begin{equation}\label{LKfull}
    \Delta \rho_{xx}=\sum _{i=\alpha,...}\sum _{p=1,...} N_{i,p}B^{1/2}R^{i,p}_{T}R^{i,p}_{\text{D}} \cos{2\pi(pF_i/B+\phi_{i,p})},
\end{equation}
where $N_{i,p}$ is oscillation amplitude, $F_{i}$ is the oscillation frequency, $\phi_{i,p}$ is the phase shift.
The temperature damping factor $R_T$ for $p=1$ is given in Eq. (\ref{tempdamp}).
The Dingle damping factor is given by
\begin{equation} \label{Dingle}
    R_{\text{D}}=\exp{\biggl(-\frac{2\pi^2k_{\text{B}}T_{\text{D}}m^*}{\hbar eB}\biggl)},
\end{equation}
where $T_{\text{D}}$ is the Dingle temperature.
The spin reduction factor $R_{\text{S}}=\cos{\bigl(\frac{p\pi g^*m^*}{2m_0}\bigl)}$ with an effective $g^*$ factor is not included because this is only applicable to the system with both time-reversal and spatial-inversion symmetries.
Phase shift $\phi$ for the $p$-th harmonics is given by
\begin{equation}
    \phi_{p}=-p/2+(\phi_{\text{B}}+\phi_{\text{R}}+\phi_{\text{Z}})/2\pi+\phi_{3\text{D}}+\phi'_{\text{Z}},
\end{equation}
where $\phi_{\text{B}}$ is the Berry phase (0 and $\pi$ for trivial and nontrivial Fermi surface, respectively) ~\cite{mikitik1999manifestation}, $\phi_{\text{R}}$ is the orbital moment, $\phi_{\text{Z}}$ is the spin Zeeman effect, and $\phi_{3\text{D}}=\pm 1/8$ for 3D Fermi surface or $\sim \pm 0$ for quasi-2D Fermi surface ($+$: hole, and $-$: electron pocket) ~\cite{shoenberg2009magnetic,li2018rules}.
$\phi'_{\text{s}}$ is the field-dependent correction related to the magnetic susceptibility ~\cite{mineev2005haas,gao2017zero,wang2019landau}.

The branches $\alpha$ and $\beta$ ($\delta$ and $\eta$) are originated from the spin-orbit splitting bands.
In such a case, $\phi_{\text{B}}$ is expected to be the nontrivial $\pi$ ~\cite{mikitik1999manifestation,xie2021kramers}, while $\phi_{\text{R}}$ and $\phi_{\text{Z}}$ average out for the orbit that encircles the TRIM ($\Gamma$ point) ~\cite{alexandradinata2018revealing,gao2017zero,fuchs2018landau}.
Although the quality of the experimental data is not sufficient to unambiguously resolve $\phi_{\text{B}}+\phi_{\text{R}}+\phi_{\text{Z}}$ by fitting, we set $\phi_{\text{B}}=\pi$ and $\phi_{\text{R}}+\phi_{\text{Z}}=0$ to reduce the number of free parameters for simplicity.
We treat $\phi'_{\text{s}}$ as a perturbation to be linear in $B$, exhibiting opposite signs between $\alpha$ and $\beta$ (and $\gamma$ and $\delta$).
This corresponds to the adiabatic limit or a weak field limit ~\cite{wang2019landau}, where the Zeeman effect is small compared to the spin-orbit splitting.
The fitting results are consistent with this approximation when the $B$ is up to 24 T.
We also note that the frequency splitting between $\alpha$ and $\beta$ remains robust regardless of the inclusion or omission of $\phi'_{\text{s}}$.
Since all the oscillation branches from $\alpha$ to $\delta$ are hole pockets with ellipsoidal shape, we set $\phi_{3\text{D}}=+1/8$ ~\cite{li2018rules}.

The best fit is given by thin curves in Fig. \ref{figAng}(a).
As the $R_{T}$ and $R_{\text{D}}$ interfere with each other, $m^*$ in $R_{T}$ and $R_{\text{D}}$ is fixed at the value obtained by the mass plot analysis.
Figure \ref{figAng}(c) shows the extracted individual oscillation components for $\alpha$ and $\beta$ branches along with their summation, and the latter well reproduces the beating of the main oscillations for both angles.
Figure \ref{figAng}(d) summarizes the angular dependence of oscillation frequencies for all the four observed branches, which is in good agreement with DFT calculations (Fig. \ref{figSBand}).

We also perform magnetization torque ($\tau$) measurements with the pulse field up to 60 T and observe de Haas-van Alphen oscillations (see Figs. \ref{figSTorque}-\ref{figSTorqueAng}).
Similar with our analysis above for the Shubnikov-de Haas oscillations, we identify four hole branches through a combination of FFT and the fitting of the LK formula to the $\tau$ vs $1/B$ curves.
The oscillation frequencies, effective mass, and angular dependence of frequencies are consistent with the analysis of the SdH oscillations (see Table \ref{table}).

The splittings of the quantum oscillations for ($\alpha$, $\beta$) and ($\gamma$, $\delta$) are a signature of SOI, which has been identified in various noncentrosymmetric systems ~\cite{onuki2022split,onuki2014chiral}.
In the approximation of the Rashba model given the polar symmetry of YAuGe, the effect of the SOI is estimated by the splitting of the band energy
\begin{equation}
\epsilon_{p\pm}=\frac{p^2}{2m^*}\mp \alpha p_{\perp},    
\end{equation}
where $p_{\perp}=\sqrt{p_x^2+p_y^2}$, $m^*$ is the effective mass, and $\alpha$ is the coefficient of SOI.
The energy splitting ($\Delta \epsilon=2\alpha p_{\text{F}}$) is related to the frequency splitting ($\Delta F$) as
\begin{equation}
    \Delta \epsilon=\frac{\hbar e \Delta F}{m^*}.
\end{equation}
The magnitude of $\Delta \epsilon$ in YAuGe is estimated to be 45 meV (for $\Delta F=F_{\beta}-F_{\alpha}$) and 27 meV (for $\Delta F=F_{\delta}-F_{\gamma}$), the order of magnitude of which is comparable with typical polar achiral rare-earth transition-metal tetrel compounds ($\Delta \epsilon\sim 10-100$ meV) ~\cite{takeuchi2006specific,okuda2007magnetic,kawai2008split,iida2011fermi}.

\subsection{Anomalous Hall effect in magnetic $R$AuGe}
A large anomalous Hall conductivity has been reported in $R$AuGe with magnetic 4f electrons at $R$ site ~\cite{ram2023multiple,kurumaji2024metamagnetism}.
The KNL in a polar system like YAuGe may be viewed as a three-dimensional generalization of the crossing point in the two-dimensional Rashba band accumulated along the $\Gamma$-A line.
The interplay between 2D Rashba models and time-reversal symmetry breaking is well studied theoretically, where time reversal symmetry breaking is expected to lift the degeneracy and generate significant Berry curvatures near the gapped crossing point ~\cite{culcer2003anomalous,dugaev2005anomalous,onoda2006intrinsic,onoda2008quantum}.
This in principle leads to concentrated Berry curvature distributions along the entire gapped nodal line, and such scenario can also be closely compared to the role of gapped nodal lines by ferromagnetic order and spin-orbit coupling near K points to anomalous Hall effect discussed for \ce{Fe3GeTe2} and \ce{Fe3Sn2}~\cite{kim2018large,fang2022ferromagnetic}.
As a last part of this paper, we discuss the connection between the anomalous transport properties of magnetic $R$AuGe and the KNLs.

A simple picture to understand the emergence of AHE in magnetic $R$AuGe is that magnetic moments break the time-reversal symmetry of the Rashba-split bands.
In the rigid band approximation, the substitution of $R$ ions with Y ion does not significantly change the electronic bands besides the exchange field from the localized 4f electrons.
This is a reasonable assumption for $R=$ Dy-Tm since the 4f electrons are deep below the Fermi level, i.e., band bending due to c-f hybridization near the Fermi level is negligible.

To analyze if these considerations are reasonable, we measure the SdH oscillation in HoAuGe up to 31 T.
As discussed in details in Appendix E, we have identified four oscillation branches with frequencies close to those observed in nonmagnetic YAuGe (see Figs. \ref{figSTempHo}, \ref{figSAngHo}).
The estimated effective mass is also comparable to those for YAuGe, indicating that the itinerant electronic bands do not strongly deviate from that of the non-magnetic YAuGe. 
Figure \ref{figBerry}(a) summarizes the angular dependence of SdH oscillation frequencies.
We note that the splitting between $\alpha$ and $\beta$ ($\gamma$ and $\eta$) in HoAuGe is enhanced from that in YAuGe, which may be due to the exchange interaction between the localized Ho 4f electrons and conduction electrons.

To capture the effect of time-reversal symmetry breaking, we calculated the band structure in ferromagnetic HoAuGe.
These calculations are done treating the 4f-shell in the open-core approximation.
Figures \ref{figBerry}(b)-(d) show the band structure in the ferromagnetic state with the magnetic moment at Ho$^{3+}$ pointing along the $c$ axis.
The degeneracy at the $\Gamma$ point protected by time-reversal symmetry in YAuGe is lifted by the magnetic moments which causes the emergence of net Berry curvature.

The anomalous Hall conductivity (AHC, $\sigma_{xy}^{\text{A}}$) as a function of $E_{\text{F}}$ is calculated for ferromagnetic states of HoAuGe and TmAuGe and summarized in Fig. \ref{figBerry}(e).
As illustrated for the case of HoAuGe in Figs. \ref{figBerry}(b)-(d), the contributions to the intrinsic anomalous Hall conductivities is concentrated near the gapped KNL.
The $E_{\text{F}}$ dependence of $\sigma_{xy}^{\text{A}}$ is shown in Fig. \ref{figBerry}(e).
Near $E_{\text{F}}$, HoAuGe has a finite positive AHC, which is in contrast to a small negative value in TmAuGe.
The observed AHC shows the $R$ dependence as it systematically decreases for $R=$ Dy, Ho, and Er and changes its sign at $R=$ Tm (Fig. \ref{figBerry}(g)).
We note, however, that the magnitude of the Hall conductivity at $E_{\text{F}}$ for both compounds is much smaller than the observed AHC.

One possibility for this discrepancy is that the dominant contribution of the measured AHC stems from extrinsic mechanisms such as skew scattering ~\cite{onoda2006intrinsic,onoda2008quantum}.
It is known that the extrinsic mechanism dominates the AHE when the system is in the clean regime, which is obtained at $E_{\text{F}}\tau/\hbar>>\pi/2$, where $\tau$ is the relaxation time.
By a rough estimate of $E_{\text{F}}(=\frac{\hbar^2 k^2{\text{F}}}{2m_{\text{eff}}})$ and $\tau(=\frac{\mu m_{\text{eff}}}{e})$, we obtain $E_{\text{F}}\tau/\hbar\sim 20-50$ for the hole pockets in HoAuGe.
This suggests that the current system may very well be on the verge of the crossover from the intrinsic to extrinsic regime, and hosts a significant extrinsic contribution to the AHE.
A second possibility is that the adopted open-core approximation is not appropriate to quantitatively describe the low-energy electronic structure.
Although we have shown that the quantum oscillation patterns and associated effective masses do not significantly change between YAuGe and HoAuGe, the Berry curvature relevant to the AHC depends strongly on the precise gap sizes induced by the long-range order in the KNLs, and these may in turn be affected by hybridization between the local moments and conduction electrons which are neglected in the open-core approximation.
Further investigation such as doping between magnetic and non-magnetic $R$AuGe and comparing their transport properties may shed additional light on the anomalous magnetotransport responses in KNL semimetals.

\section{Conclusion}
In conclusion, we have revealed the electronic structure of a KNL semimetal candidate YAuGe, and established the existence of KNLs by symmetry considerations and DFT calculations.
The high-mobile hole carriers originated from KNLs crossing the Fermi level are identified by ARPES and quantum oscillations.
Additionally, the spin splitting of these bands—a hallmark of inversion symmetry breaking in the lattice—has been resolved through quantum oscillation studies.
We also discussed the anomalous magnetotransport properties observed in $R$AuGe, suggesting that both intrinsic and extrinsic mechanisms may contribute, with the intrinsic effects arising from the interplay between time-reversal symmetry breaking and the KNLs.
These insights highlight the potential of KNL semimetals as platforms for exploring spin-orbit coupling-induced spin splitting and enhanced anomalous magnetotransport properties when broken time-reversal symmetry is incorporated.

\section{Acknowledgements}
T.K. was supported by Ministry of Education Culture Sports Science and Technology (MEXT) Leading Initiative for Excellent Young Researchers (JPMXS0320200135) and Inamori Foundation.
This study was partially supported by Japan Society for the Promotion of Science (JSPS) KAKENHI Grant-in-Aid (No. 21K13874, 23K13068, JP19H05826, 19H01835, 22H00109, 22H04933, 23K22447, 21H05470),  Institute of Quantum Information and Matter, an NSF Physics Frontier Center via PHY- 2317110 and Gordon and Betty Moore Foundation via GBMF12765.
This work was partly performed under the GIMRT Program of the Institute for Materials Research, Tohoku University (Proposal No. 202203-HMKPA-0064).
The experiment at National High Magnetic Field Laboratory (NHMFL), which is supported by National Science Foundation Cooperative Agreement No. DMR-2128556 and the State of Florida.
The synchrotron ARPES experiments are performed under the approval of the Photon Factory Program Advisory Committee (Proposal No. 2021G141) and the Japan Synchrotron Radiation Research Institute (JASRI) (Proposal No. 2022B1332).
The magnetization torque measurements were carried out by the joint research in the Institute for Solid State Physics, the University of Tokyo (No.202209-HMBXX-0088).
This work was partly performed using the facilities of the Materials Design and Characterization Laboratory in the Institute for Solid State Physics, the University of Tokyo.

\bibliography{reference}

\clearpage
\appendix
\gdef\thefigure{\thesection.\arabic{figure}}    
\gdef\thetable{\thesection.\arabic{table}}
\setcounter{figure}{0}
\setcounter{table}{0}
\section{Methods}
Single crystals were grown by using Au-Ge self flux as reported in Ref. \onlinecite{kurumaji2023single}.
Electrical transport measurements were performed by a conventional five probe method at typical frequency near 37 Hz.
The transport properties at low temperatures in a magnetic field were measured using a commercial superconducting magnet and cryostat.
Magnetic torque was measured by using a piezoresistive microcantilever ~\cite{ohmichi2002torque} in pulse magnetic fields up to 60 T at ISSP.
Shubnikov-de Haas oscillations in the longitudinal resistivity ($\rho_{xx}$) in YAuGe and HoAuGe were measured at IMS Tohoku university (up to 24 T) in Japan and NHMFL (up to 32 T) in Florida, respectively.
Specific heat was measured using a commercial system (heat capacity option of a Quantum Design Physical Property Measurement System).

We perform fully-relativistic density-functional theory (DFT) calculations based on the experimental crystal structure with the code FPLO v22.01-63 ~\cite{Koepernik1999} and the generalized gradient approximation (GGA)~\cite{Perdew1997}.
We use the enhanced basis setup, as described in Ref. \onlinecite{lejaeghere2016reproducibility} and for HoAuGe and TmAuGe we use the open-core approximation for the 4f levels, fixing the magnetic moment to the values expected for trivalent ions.
BZ integrations were performed with a tetrahedron method based on a $k$-mesh having $24\times 24\times 12$ subdivisions.
For calculation of the AHC, we construct maximally-projected Wannier functions with pyfplo ~\cite{PhysRevB.107.235135}.
The Wannier models typically include orbitals associated with Au 5d, 6s and 6p; Ge 4s and 4p as well as $R$ 4d.

Vacuum-ultraviolet and soft x-ray ARPES measurements were performed at BL28A in Photon Factory (KEK)~\cite{kitamura2022development} and BL25SU in SPring-8~\cite{senba2016upgrade}, respectively, both equipped micro-focused beams with Scienta Omicron DA30 electron analyzers.
Single-crystalline YAuGe were cleaved $\it{in}$ $\it{situ}$ by standard top-post method along the (001) plane at around 20 K.
The vacuum level was kept better than $1\times 10^{-10}$ Torr throughout the measurements.
The total energy resolution of vacuum-ultraviolet ARPES (presented in Fig. \ref{figARPES}) and soft x-ray ARPES (presented in Appendix C) was set to 35 meV and 40-80 meV, respectively.

\section{Symmetry analysis and density functional theory calculations of the electronic structure of YAuGe}
The crystal structure of YAuGe belongs to the nonsymmorphic noncentrosymmetric space group $P6_3mc$ (SG.186, $C^4_{6v}$).
The primitive lattice vectors are $\bm{a}_1=(0,-a,0)$, $\bm{a}_2=(\frac{\sqrt{3}a}{2},\frac{a}{2},0)$, and $\bm{a}_3=(0,0,c)$, where the lattice constants are $a=4.41$ \AA, $c=7.27$ \AA, respectively, at $T=30$ K ~\cite{kurumaji2024metamagnetism}.
In a unit cell, there are two sets of formula unit, i.e., $Z=2$ (see Fig. \ref{figBand}).
The reciprocal lattice vectors are $\bm{b}_1=\frac{2\pi}{a}(\frac{1}{\sqrt{3}},-1,0)$, $\bm{b}_2=\frac{2\pi}{a}(\frac{2}{\sqrt{3}},0,0)$, and $\bm{b}_3=\frac{2\pi}{c}(0,0,1)$.

\begin{figure}[t]
	\includegraphics[width =  0.4\columnwidth]{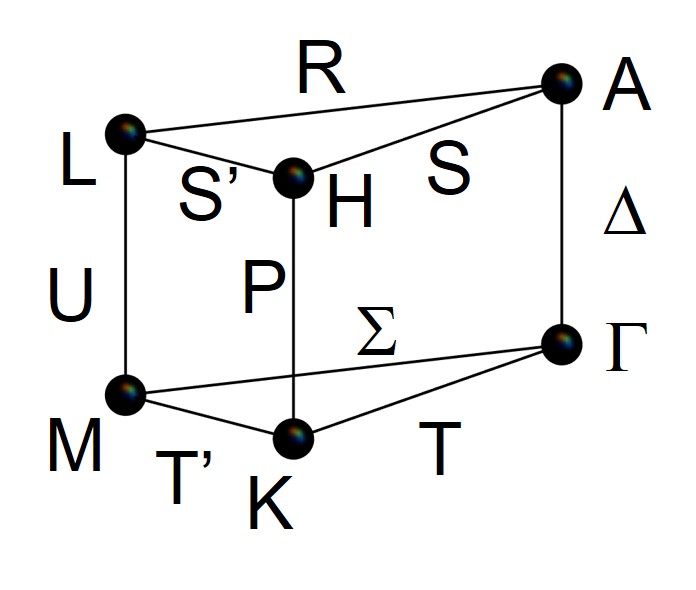}
	\caption{\label{figBZ}Symbols of the high-symmetry points and lines in the BZ of YAuGe.
 }
\end{figure}

The degeneracy of electronic bands at high-symmetry points and along lines in BZ (Fig. \ref{figBZ}) can be systematically analyzed by group theory ~\cite{bradley2010mathematical}.
Tables \ref{tableSIR} and \ref{tableDIR} summarize the Herring's little groups $^{\text{H}}G_{\bm{k}}(=G_{\bm{k}}/T_{\bm{k}})$ ($G_{\bm{k}}$ is the little group at momentum $\bm{k}$ and $T_{\bm{k}}$ is the group of translational symmetry operations $\{E|\bm{t}\}$ with $\exp{(-\text{i}\bm{k}\cdot \bm{t})}=1$) for $P6_3mc$ and their single and double-valued irreducible representations (IRs).

TRIMs are located at $\Gamma$ ($=(0,0,0)$), A ($=(0,0,\frac{1}{2})$), M ($=(0,\frac{1}{2},0)$), and L ($=(0,\frac{1}{2},\frac{1}{2})$) points, where the coordinates are taken by the reciprcal lattice vectors.
The $\Gamma$ALM-plane is invariant under a mirror inversion symmetry ($s_0=\{m_{100}|000\}$, where the coordinates are taken by the primitive lattice vectors).
$\Delta^x$ and $U^x$ lines connecting $\Gamma$-A and M-L, respectively, are two-fold degenerate even with the SOI, which are identified as KNL.
We note that A and L are four-fold degenerate even with the SOI.
This is owing to $6_3$-screw ($s_1=\{C^+_6|00\frac{1}{2}\}$) and $c$-glide ($s_2=\{m_{1\bar{1}0}|00\frac{1}{2}\}$) symmetries, respectively, of the lattice as pointed out in Ref. \onlinecite{zhang2018topological}.
Furthermore, the $6_3$-screw makes the $k_z=\pi/c$-plane at the BZ boundary a nodal plane (NP) because the points upon this plane are invariant under the combination of time-reversal ($\mathcal{T}$) and the $6_3$ screw ($s_1\mathcal{T}$) ~\cite{chang2018topological}.
We note that $P^x$ (K-H) may have two-fold degenerate bands when their IR belong to $\bar{E}_1(2)$ (see Table~\ref{tableDIR}), while this is beyond the scope of this study as the band energy at K-H line in YAuGe is sufficiently away from $E_{\text{F}}$.

\begin{table*}
\caption{\label{tableSIR} The Herring's little group $^{\text{H}}G^{\bm{k}}=G^{\bm{k}}/T^{\bm{k}}$ on points or lines of symmetry in the BZ of YAuGe, and the single-valued irreducible representations (IRs). 
The number in the parenthesis following each IR represents the dimension of the corresponding IR.
The label $\leftrightarrow$ between two IRs means that these two IRs are paired up to form time-reversal-invariant representations (co-representations).
The total number of degeneracy is obtained by doubling the number in parentheses, which is due to the spin degrees of freedom.
The label convention is adopted as the same of Ref. \onlinecite{bradley2010mathematical}.
}
\begin{tabular}{ccc}
\hline
 & $^{\text{H}}G^{\bm{k}}$  & Single-valued IRs \\
\hline
\hline
Point of symmetry &  &  \\
$\Gamma$ & $G^{3}_{12}$ & $A_1(1)$, $A_2(1)$, $B_2(1)$, $B_1(1)$, $E_2(2)$, $E_1(2)$ \\
M & $G^{2}_{4}\otimes T_2$ & $A_1(1)$, $A_2(1)$, $B_1(1)$, $B_2(1)$ \\
A & $G^{4}_{24}$ & $^1E_1(1)$$\leftrightarrow$$^2E_1(1)$, $^1E_2(1)$$\leftrightarrow$$^2E_2(1)$, $^1F(2)$$\leftrightarrow$$^2F(2)$ \\
L & $G^{2}_{8}$ & $^2E'(1)$$\leftrightarrow$$^1E'(1)$, $^2E''(1)$$\leftrightarrow$$^1E''(1)$\\
K & $G^{2}_{6}\otimes T_3$ & $A_1(1)$, $A_2(1)$, $E(2)$ \\
H & $G^{4}_{12}\otimes T_3\otimes T_2$ & $^1E_1(1)$$\leftrightarrow$$^2E_1(1)$, $E_2(2)$ \\
\hline
Line of symmetry &  &  \\
$\Delta^x$ & $G^{3}_{12}$ & $A_1(1)$, $A_2(1)$, $B_2(1)$, $B_1(1)$, $E_2(2)$, $E_1(2)$ \\
$U^x$ & $G^{2}_{4}$ & $A_1(1)$, $A_2(1)$, $B_1(1)$, $B_2(1)$ \\
$P^x$ & $G^{2}_{6}$ & $A_1(1)$, $A_2(1)$, $E(2)$ \\
$T^x$ & $G^{1}_{2}$ & $A'(1)$, $A''(1)$ \\
$S^x$ & $G^{1}_{2}$ & $A'(1)$$\leftrightarrow$$A''(1)$ \\
$T'^x$ & $G^{1}_{2}$ & $A'(1)$, $A''(1)$ \\
$S'^x$ & $G^{1}_{2}$ & $A'(1)$$\leftrightarrow$$A''(1)$ \\
$\Sigma ^x$ & $G^{1}_{2}$ & $A'(1)$, $A''(1)$ \\
$R^x$ & $G^{1}_{2}$ & $A'(1)$($\leftrightarrow$$A'(1)$), $A''(1)$($\leftrightarrow$$A''(1)$) \\
\hline
\end{tabular}
\end{table*}

\begin{table*}
\caption{\label{tableDIR} The double group and double-valued irreducible representations of the quotient group $^{\text{H}}G^{\bm{k}}=G^{\bm{k}}/T^{\bm{k}}$ on points or lines of symmetry in the BZ of YAuGe.
The label convention is adopted as the same of Ref. \onlinecite{bradley2010mathematical}.
}
\begin{tabular}{ccc}
\hline
 & Double group of $^{\text{H}}G^{\bm{k}}$  & Double-valued IRs \\
\hline
\hline
Point of symmetry &  &  \\
$\Gamma$ & $G^{11}_{24}$ & $\bar{E}_1(2)$, $\bar{E}_2(2)$, $\bar{E}_3(2)$ \\
M & $G^{5}_{8}\otimes T_2$ & $\bar{E}(2)$ \\
A & $G^{13}_{48}$ & $\bar{E}(2)$($\leftrightarrow$$\bar{E}(2)$), $^1\bar{F}(2)$$\leftrightarrow$$^2\bar{F}(2)$ \\
L & $G^{8}_{16}$ & $\bar{E}(2)$($\leftrightarrow$$\bar{E}(2)$)\\
K & $G^{4}_{12}\otimes T_3$ & $^1\bar{E}(1)$, $^2\bar{E}(1)$, $\bar{E}_1(2)$ \\
H & $G^{4}_{12}\otimes T_3\otimes T_2$ & $\bar{A}_1(1)$$\leftrightarrow$$\bar{A}_2(1)$, $\bar{E}(2)$ \\
\hline
Line of symmetry &  &  \\
$\Delta^x$ & $G^{11}_{24}$ & $\bar{E}_1(2)$, $\bar{E}_2(2)$, $\bar{E}_3(2)$ \\
$U^x$ & $G^{5}_{8}$ & $\bar{E}(2)$ \\
$P^x$ & $G^{4}_{12}$ & $^1\bar{E}(1)$, $^2\bar{E}(1)$, $\bar{E}_1(2)$ \\
$T^x$ & $G^{1}_{4}$ & $^2\bar{E}(1)$,$^1\bar{E}(1)$ \\
$S^x$ & $G^{1}_{4}$ & $^2\bar{E}(1)$$\leftrightarrow$$^1\bar{E}(1)$ \\
$T'^x$ & $G^{1}_{4}$ & $^2\bar{E}(1)$,$^1\bar{E}(1)$ \\
$S'^x$ & $G^{1}_{4}$ & $^2\bar{E}(1)$$\leftrightarrow$$^1\bar{E}(1)$ \\
$\Sigma ^x$ & $G^{1}_{4}$ & $^2\bar{E}(1)$,$^1\bar{E}(1)$ \\
$R^x$ & $G^{1}_{4}$ & $^2\bar{E}(1)$($\leftrightarrow$$^2\bar{E}(1)$),$^1\bar{E}(1)$($\leftrightarrow$$^1\bar{E}(1)$) \\
\hline
\end{tabular}
\end{table*}

The symmetry analysis above is consistent with the DFT calculation shown in Figs. \ref{figBand}(d)-(e).
In the presence of SOI (Fig. \ref{figBand}(d)), the band energy has four-fold degeneracy at A and L points, and they split into two two-fold degenerate bands along $\Delta^x$ ($\Gamma$-A line), $R^x$ (A-L line), $U^x$ (M-L line), $S^x$ (A-H line), and $S'^x$ (L-H line) ~\cite{wu2021symmetry}.
Along $\Sigma^x$ ($\Gamma$-M line) and $T^x$ ($\Gamma$-K line), all the bands lift the degeneracy, which ensure the splitting of the bands from $\alpha$ to $\eta$ at Fermi energy ($E_{\text{F}}$) in $k_xk_y$ plane.
Along the $k_z$ direction, pairs of Fermi surfaces, i.e, $\alpha$-$\beta$, $\gamma$-$\delta$, and $\zeta$-$\eta$, has a touching point due to the KNL nature.
We note that the Fermi surface of $\zeta$-$\eta$ pair has other touching points as denoted in Fig. \ref{figBand}(i).
By shifting $E_{\text{F}}$, the touching points draw loops in the BZ as shown in Fig. \ref{figBand}(l), which are pinned on the $\Gamma$ALM plane by the $s_0$ symmetry ~\cite{xie2021kramers}.

\clearpage
\section{Soft x-ray ARPES experiments in YAuGe}
\begin{figure}[t]
	\includegraphics[width =  \columnwidth]{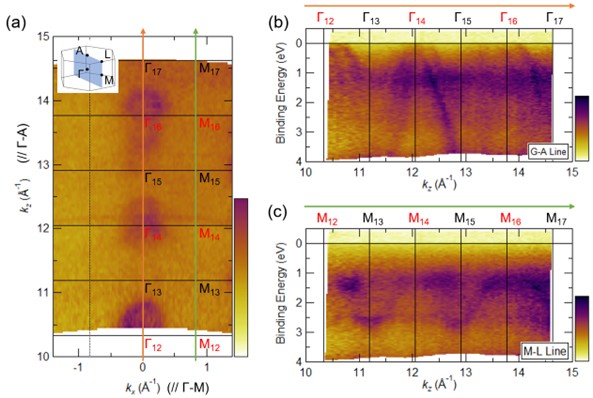}
	\caption{\label{figSsoft}(a) ARPES intensity plots at $E_{\text{F}}$ [integral width: 50 meV] in the $\Gamma$-M-L-A plane for YAuGe, collected at $T=20$ K using $E_{\text{i}}=400$-800 eV incident photons.
    (b), (c) ARPES images along $\Gamma$-A and M-L lines (see orange/green lines in (a)).
    }
\end{figure}

To investigate the three-dimensionality of band structures in YAuGe, we have performed photon-energy-dependent ARPES measurements in the soft x-ray photon energy regime.
Figure \ref{figSsoft}(a) displays the ARPES intensity plots at $E_{\text{F}}$ in the $\Gamma$-M-L-A plane (see inset), obtained by changing the incident photons from $E_{\text{i}}=400$ to 800 eV with interval of 5 eV.
For this soft x-ray dataset, we assume the inner potential of 20 eV to plausibly convert the photoelectron emission angle into momentum.
We find the signatures of ellipse Fermi pockets centered at every even $\Gamma$ points (such as $\Gamma_{14}$ and $\Gamma_{16}$), but such intensity is mostly absent at odd  $\Gamma$ ($\Gamma_{13}$ and $\Gamma_{15}$).
This double periodicity (i.e., $4\pi/c$ periodicity) of ARPES spectra can also be observed in the band dispersion along $\Gamma$-A and M-L directions as shown in Fig. \ref{figSsoft}(b) and (c).
Such feature is characteristic to materials with a nonsymmorphic space group, as known for example in 2$H$-WSe$_2$~\cite{finteis1997occupied} and BiTeCl~\cite{landolt2013bulk}.

\section{Quantum oscillations and specific heat in YAuGe}
\subsection{Shubnikov-de Haas oscillations}
\begin{figure}[t]
	\includegraphics[width =  0.8\columnwidth]{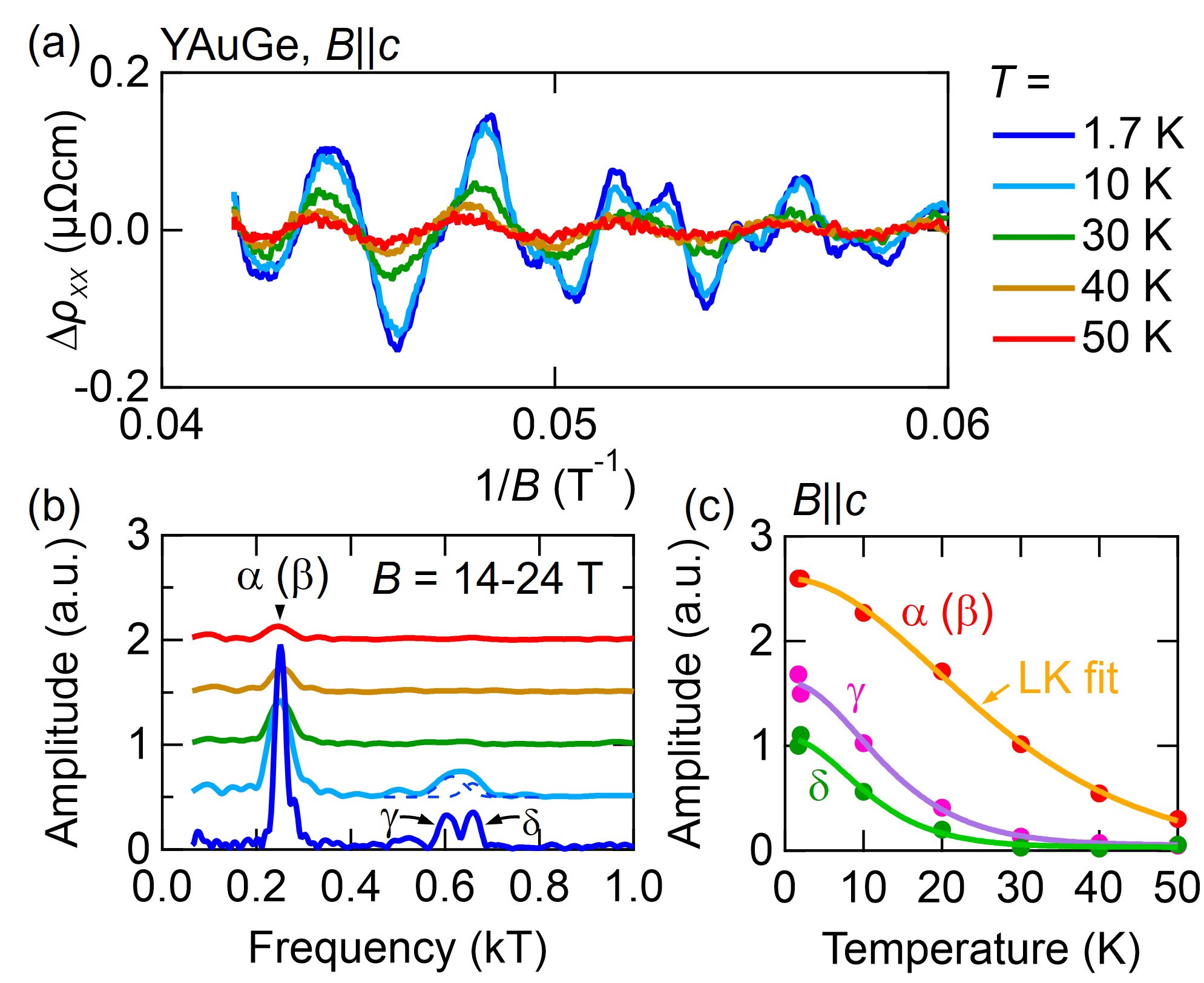}
	\caption{\label{figTemp} (a) $B$-inverse field dependence of the background-subtracted resistivity ($\Delta \rho_{xx}$) measured at various temperatures with the $B$ along the $c$ axis.
 (b) FFT of SdH oscillations for $B\parallel c$ at various temperatures.
 Positions of the branches $\alpha$, $\gamma$, and $\delta$ are assigned.
 The $\beta$ branch is overlapped with $\alpha$.
 Dashed curve is the gaussian functions that fit a broad peak for $\gamma$ and $\delta$ pockets.
 (c) Temperature dependence of peak amplitude (closed circles) and fits with LK formula (solid lines) for each branch.
 }
\end{figure}

To characterize the effective mass of each branch, we measure the SdH oscillations in YAuGe at various temperatures with $B\parallel c$ (Fig. \ref{figTemp}(a)).
As shown in Fig. \ref{figTemp}(b), the FFT identifies the three branches, $\alpha$, $\gamma$, and $\delta$ at the lowest temperatures, while the frequency for the $\beta$ branch is overlapped with that of $\alpha$.
Increasing the temperature, the frequency peaks for $\gamma$ and $\delta$ merge to a single peak.
The $\gamma$ and $\delta$ are resolved by a fit of the peak with two Gaussian functions at fixed frequencies observed at $T=1.7$ K as shown by blue dashed curve for $T=10$ K.
The temperature dependence of each oscillation amplitude is shown in Fig. \ref{figTemp}(c).
From these data we estimate the effective mass by using the temperature damping factor in a Lifshitz-Kosevich (LK) formula as below ~\cite{shoenberg2009magnetic}.
\begin{equation}\label{tempdamp}
    R_{T}=\frac{2\pi^2 p k_{\text{B}}Tm^*}{\hbar eB}\sinh^{-1}\biggl( \frac{2\pi^2 p k_{\text{B}}Tm^*}{\hbar eB}\biggl),
\end{equation}
where $\hbar$ is the Planck constant divided by $2\pi$, $k_{\text{B}}$ is the Boltzmann constant, $e(>0)$ is the elementary charge, $m^*$ is the effective mass, $p$ is the number of harmonics ~\cite{shoenberg2009magnetic}.
$B$ is replaced with the average $B_{\text{av}}=\{\frac{1}{2}(1/B_{\text{h}}+1/B_{\text{l}})\}^{-1}$ between the highest ($B_{\text{h}}=24$ T) and lowest ($B_{\text{l}}=14$ T) fields.
The physical parameters of each branch are summarized in Table \ref{table}.

\begin{figure}[t]
	\includegraphics[width =  \columnwidth]{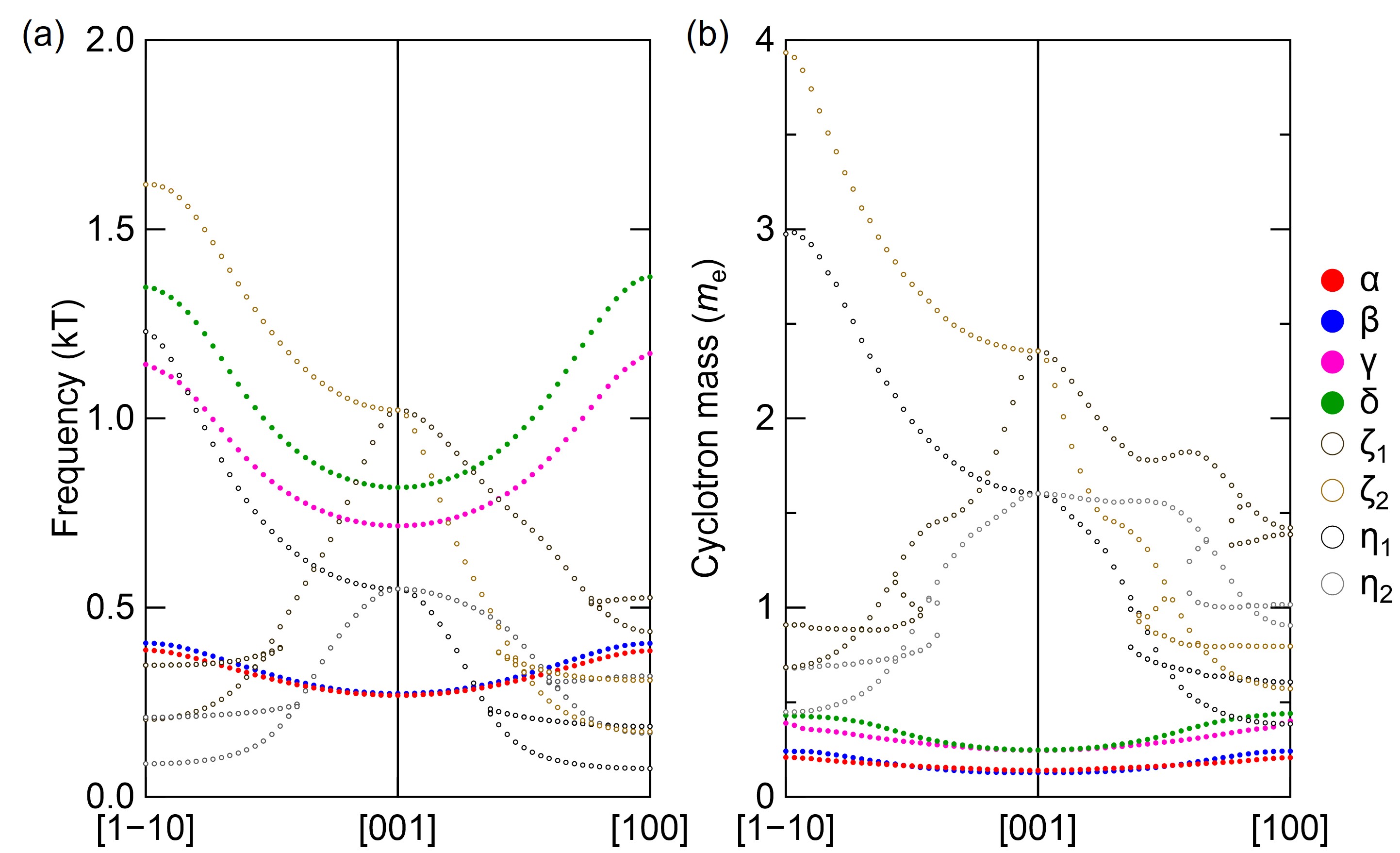}
	\caption{\label{figSBand} The DFT calculations of (a)-(b) magnetic field orientation dependence of (a) quantum oscillation frequency and (b) effective mass for each branch.
}
\end{figure}

The angular dependence of oscillation frequency and the effective mass is simulated by DFT calculations.
Figure \ref{figSBand}(a) shows the frequencies of FS extrema for the plane perpendicular to the magnetic field.
The frequencies for the hole pockets become higher as the field is tilted towards the $ab$ plane.
This is due to the elliptically elongeted FSs along the $k_z$ directions (see Figs. \ref{figBand}(g)-(h)).
Figure \ref{figSBand}(b) is the angular dependence of the effective mass for each frequency branch.
Those for the hole pockets are $0.1\sim 0.2m_e$, while the electron bands have relatively heavy mass.

\subsection{Specific heat}
\begin{figure}[t]
	\includegraphics[width =  0.6\columnwidth]{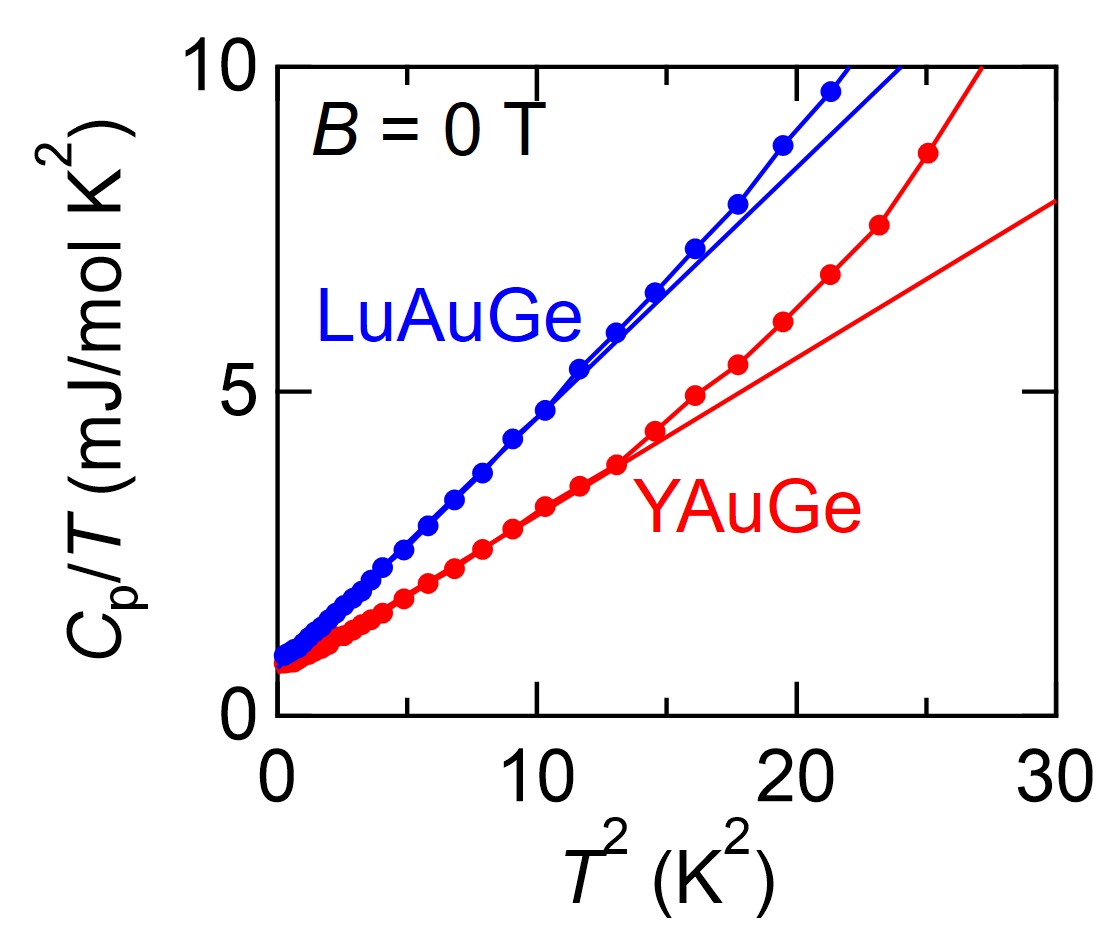}
	\caption{\label{figDoS} Temperature dependence of zero field specific heat ($C_{\text{p}}$) for YAuGe and LuAuGe.
 Solid lines are the linear fit for the low-temperature slope.
}
\end{figure}

To see the signature of the electron pockets, we have measured the temperature dependence of specific heat in YAuGe down to 0.5 K.
As shown in Fig. \ref{figDoS}, $C_{\text{p}}/T$ vs. $T^2$ plot has a linear slope, which can be fit with
\begin{equation}
    C_{\text{p}}/T=\gamma_{\text{c}}+\beta_{\text{ph}} T^2,
\end{equation}
where $\gamma_{\text{c}}$ and $\beta_{\text{ph}}$ is the coefficient for conduction electrons and phonons, respectively.
As a comparison, we also measured $C_{\text{p}}$ for LuAuGe, where $\gamma_{\text{c}}$ is almost identical, which is reasonable for the nonmagnetic sibling (isoelectronic and isostructural) compounds.
The steeper $T^2$-evolution in LuAuGe, i.e., higher $\beta_{\text{ph}}$, is consistent with the Debye model, giving a lower Debye temperature due to the heavier molecular masses in LuAuGe than in YAuGe ~\cite{bouvier1991specific,avila2004anisotropic}.
The Debye temperature ($\Theta_{\text{D}}$) is obtained by $\Theta_{\text{D}}=(\frac{12\pi^4k_{\text{B}}N_{\text{A}}n}{5\beta_{\text{ph}}})^{1/3}$, where $n$ ($=3$) is the number of atoms in a formula unit.
We obtain $\Theta_{\text{D,Y}}=289$ K and $\Theta_{\text{D,Lu}}=247$ K.
The coefficient $\gamma_{\text{c}}$ is proportional to the density of states ($D(E_{\text{F}})$) as
\begin{equation}
    \gamma_{\text{c}}=\frac{\pi^2}{3}k_{\text{B}}^2D(E_{\text{F}}).
\end{equation}
The observed $\gamma_{\text{c}}$ ($=0.659$ mJ/mol K$^2$ in YAuGe, and $0.741$ mJ/mol K$^2$ in LuAuGe) correspond to $D_{\text{tot}}=0.280$ eV$^{-1}$ f.u.$^{-1}$ and 0.314 eV$^{-1}$ f.u.$^{-1}$, respectively

The contribution from the electron pockets ($D_{\text{e}}$) in YAuGe can be obtained by subtracting the hole components ($D_{\text{h}}$) from $D_{\text{tot}}$.
We estimate the $D_{\text{h}}$ from the SdH results, by assuming that the shape of the Fermi surfaces is elliptical with the long axis ($2k_{\text{F}\parallel}$) along the $c^*$ axis and short axis ($2k_{\text{F}\perp}$) in the $ab$ plane.
This approximation is in good agreement with the angular dependence of the SdH oscillations (Fig. \ref{figAng}(d)).
The density of states ($D_{\text{ellip}}$) is given by
\begin{equation}
    D_{\text{ellip}}=\frac{1}{8\pi^3}\frac{\partial}{\partial E_{\text{F}}}(\frac{4\pi}{3}k_{\text{F}\parallel}k^2_{\text{F}\perp})=\frac{1}{6\pi^2\hbar^2}(2k_{\text{F}\parallel}m^*_{\perp}+\frac{k^2_{\text{F}\perp}m^*_{\parallel}}{k_{\text{F}\parallel}}),
\end{equation}
where $m^*_{\parallel,\perp}$ is the effective mass along $c^*$ axis and the $ab$ plane.
By assuming $m^*_{\parallel,\perp}$ to be identical to the estimated effective masses for each band in $B\parallel c$ (Fig. \ref{figTemp}(c)), we obtain $D_{\text{h}}=D_{\alpha}+D_{\beta}+D_{\gamma}+D_{\delta}=0.041$ eV$^{-1}$ f.u.$^{-1}$.
As a consequence, we obtain the $D_{\text{e}}=D_{\text{tot}}-D_{\text{h}}=3(D_{\zeta}+D_{\eta})=0.24$ eV$^{-1}$ f.u.$^{-1}$.
The factor three represents the valley degrees of freedom for the pocket at the M point.
The dominant electron contribution to $D(E_{\text{F}})$ is consistent with the heavy effective mass of electron pockets.
By using the relationship between $D_{\text{e}}$ and the effective mass: $D_{\text{e}}=6\frac{k_{\text{F}}m^*_{\zeta,\eta}}{\pi^2 \hbar^2 }$ for a spherical FS, we obtain the rough estimate of $m^*_{\zeta,\eta}$ for the electron bands, corresponding to $m^*_{\zeta,\eta}=0.4m_0$, where the factor six is the number of electron pockets in the BZ, and we used the carrier density of electrons $n_{\text{e}}\sim 1.77 \times 10^{20}$ cm$^{-3}$ obtained in Ref. \onlinecite{kurumaji2024metamagnetism} to estimate $k_{\text{F}}$.
We note that the effective mass is heavier than those for the hole pockets but underestimated compared to the DFT calculations (Fig. \ref{figSBand}), which stems from the fact that the electron pockets, $\zeta$ and $\eta$, are anisotropic in shape (Fig. \ref{figBand}(i)), and the effective cyclotron mass is also anisotropic with respect to the applied magnetic field (Fig. \ref{figSBand}(b)).

\subsection{de Haas-van Alphen oscillations}
\begin{figure}[t]
	\includegraphics[width =  \columnwidth]{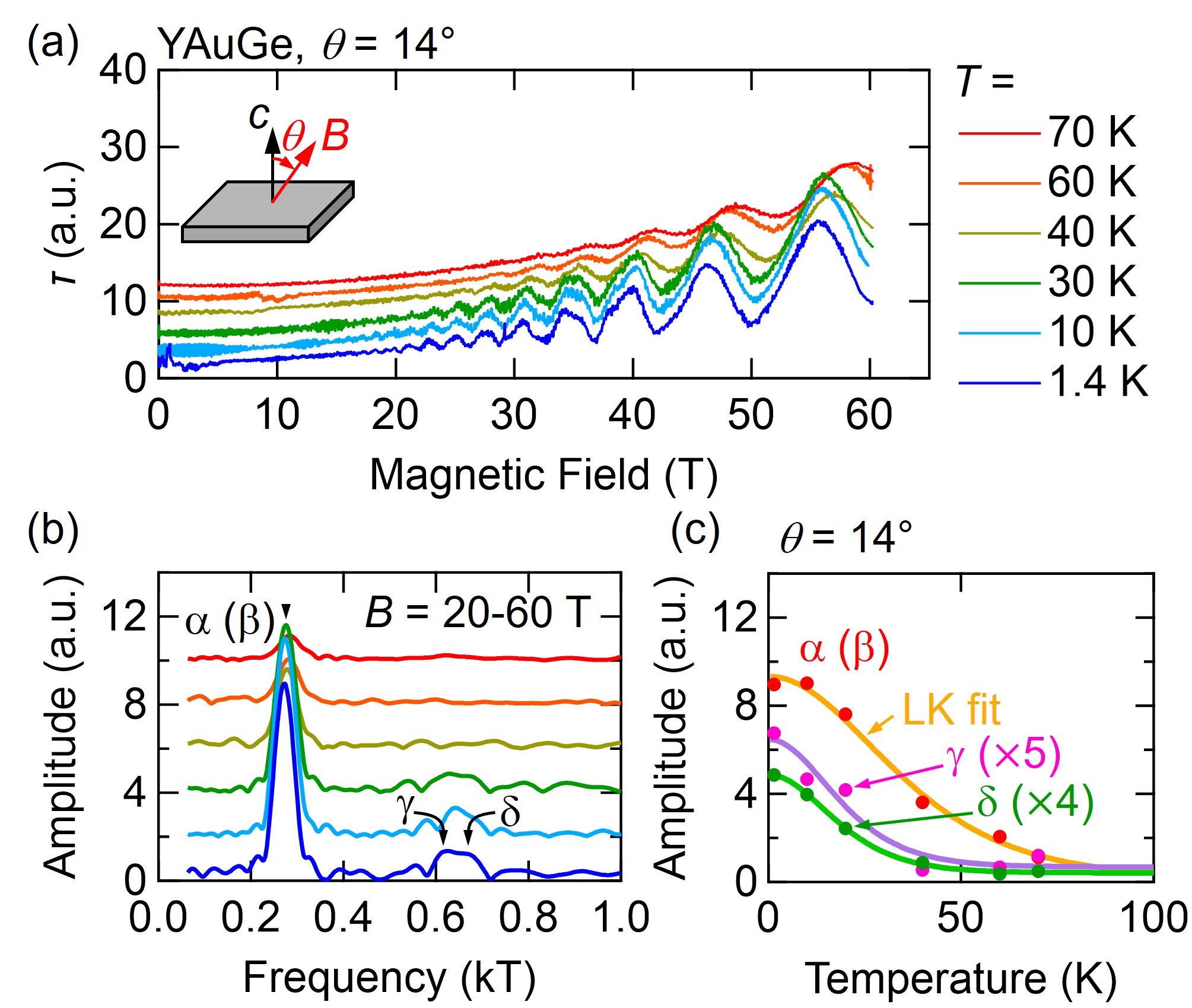}
	\caption{\label{figSTorque}(a) Field-dependence of magnetization torque ($\tau$) at various temperatures with a magnetic field tilted from the $c$ axis by $\theta = 14^{\circ}$.
 Curves for $T>1.4$ K are shifted for visibility.
(b) FFT of dHvA oscillations of $\tau$ in the field range $B=20-60$ T.
(c) Temperature dependence of peak amplitude (closed circles) and fits with LK formula (solid lines) for each branch.
}
\end{figure}

We also observe the de Haas-van Alphen oscillations in the magnetization torque ($\tau$).
Figure \ref{figSTorque}(a) shows the field dependence of $\tau$ at various temperatures, where the $B$ is canted by $\theta = 14^{\circ}$ from the $c$ axis.
The FFT of the background subtracted component $\Delta\tau$ is shown in Fig. \ref{figSTorque}(b).
As in the case of the SdH oscillation, $\alpha$ and $\beta$ branches are hard to be resolved, while $\gamma$ and $\delta$ branches can be fitted with two Gaussian functions.
The temperature dependence of each oscillation amplitude (Fig. \ref{figSTorque}(c)) is analyzed by Eq. (\ref{tempdamp}), and the obtained physical parameters such as the effective mass are summarized in Table \ref{table}.

\begin{figure}[t]
	\includegraphics[width =  \columnwidth]{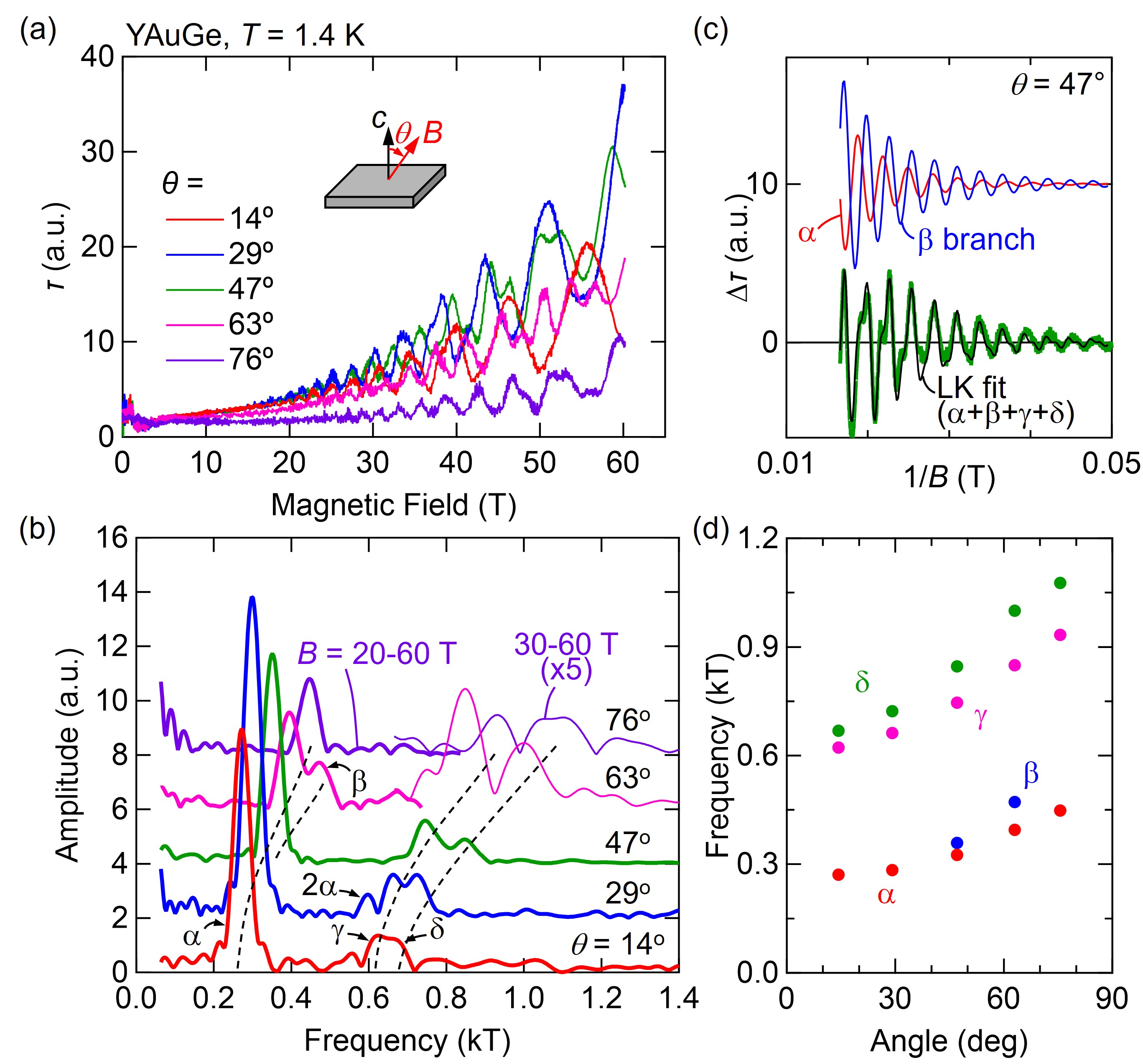}
	\caption{\label{figSTorqueAng}(a) Field dependence of $\tau$ with various $\theta$ at $T=1.4$ K.
(b) FFT of dHvA oscillations in the field range $B=20-60$ T (thick) and $B=30-60$ T (thin).
Dashed curve is guide to the eye.
(c) Comparison between dHvA oscillations at $\theta=47^{\circ}$ and the LK fit.
 Thin black curve is the fitting result expressed by the summation of oscillation components for $\alpha$, $\beta$, $\gamma$, and $\delta$.
 Thin red (blue) curve is the oscillation component associated to $\alpha$ ($\beta$).
(d) Angular ($\theta$) dependence of the dHvA oscillation frequency for each branch (closed circles).
}
\end{figure}

The angular ($\theta$) dependence of the de Haas-van Alphen oscillations is shown in Fig. \ref{figSTorqueAng}(a).
The FFT in Fig. \ref{figSTorqueAng}(b) clearly resolve the branches $\gamma$ and $\delta$ at higher angle.
As for the $\alpha$ and $\beta$ branches, we observe their splitting at $\theta = 63^{\circ}$.
The oscillation profile for $\theta=47^{\circ}$ (Fig. \ref{figSTorqueAng}(c)) is nonmonotonic in $1/B$, suggesting a beating between $\alpha$ and $\beta$ branches as observed in SdH oscillations (Fig. \ref{figAng}(a)).
To further resolve the oscillations, we fit the raw data with the LK formula, where the oscillation part ($\tau_{\text{osc}}$) of the magnetization torque is given below ~\cite{shoenberg2009magnetic}.
\begin{equation}
\tau_{\text{osc}}=\sum _{i=\alpha,...}\sum _{p=1,...} N_{i,p}B^{3/2}R^{i,p}_{T}R^{i,p}_{\text{D}} \sin{2\pi(pF_i/B+\phi_{i,p})},
\end{equation}
The extracted oscillation patterns for the $\alpha$ and $\beta$ branches are shown in Fig. \ref{figSTorqueAng}(c).
The summation of those well reproduces the amplitude modulation of the main oscillation in the raw data.
The angular dependence of the oscillation frequenceis is plotted in Fig. \ref{figSTorqueAng}(d), which is in good agreement with SdH oscillations.
The FS parameters are summarized in Table \ref{table}.

\section{Shubnikov-de Haas oscillation in HoAuGe and anomalous Hall effect in $R$AuGe}
We have performed SdH oscillation measurement in HoAuGe by using the DC magnet in NHMFL, which can apply a high magnetic field up to 31 T.
In magnetic materials with magnetization $M$, the quantum oscillations are known to be periodic to $1/B$ with $B=\mu_0H_{\text{ext}}-N_{\text{D}}M+M$, where $\mu_0H_{\text{ext}}$ is the external magnetic field and $N_{\text{D}}$ is the demagentization factor ~\cite{anderson1963haas}.
In HoAuGe, the difference between $\mu_0H_{\text{ext}}$ and $B$ is around 0.7 T when the saturation moment $10\mu_{\text{B}}$ of a free Ho$^{3+}$ ion is along the $c$ axis.
Due to the lack of knowledge on the magnetization curve above 7 T, an accurate $B$ vs. $\mu_0H_{\text{ext}}$ curve is absent.
The data is analyzed with $\mu_0H_{\text{ext}}$ instead of $B$.
The oscillation frequencies are, thus, underestimated by $\sim5$\% in maximum.

\begin{figure}[t]
	\includegraphics[width =  0.8\columnwidth]{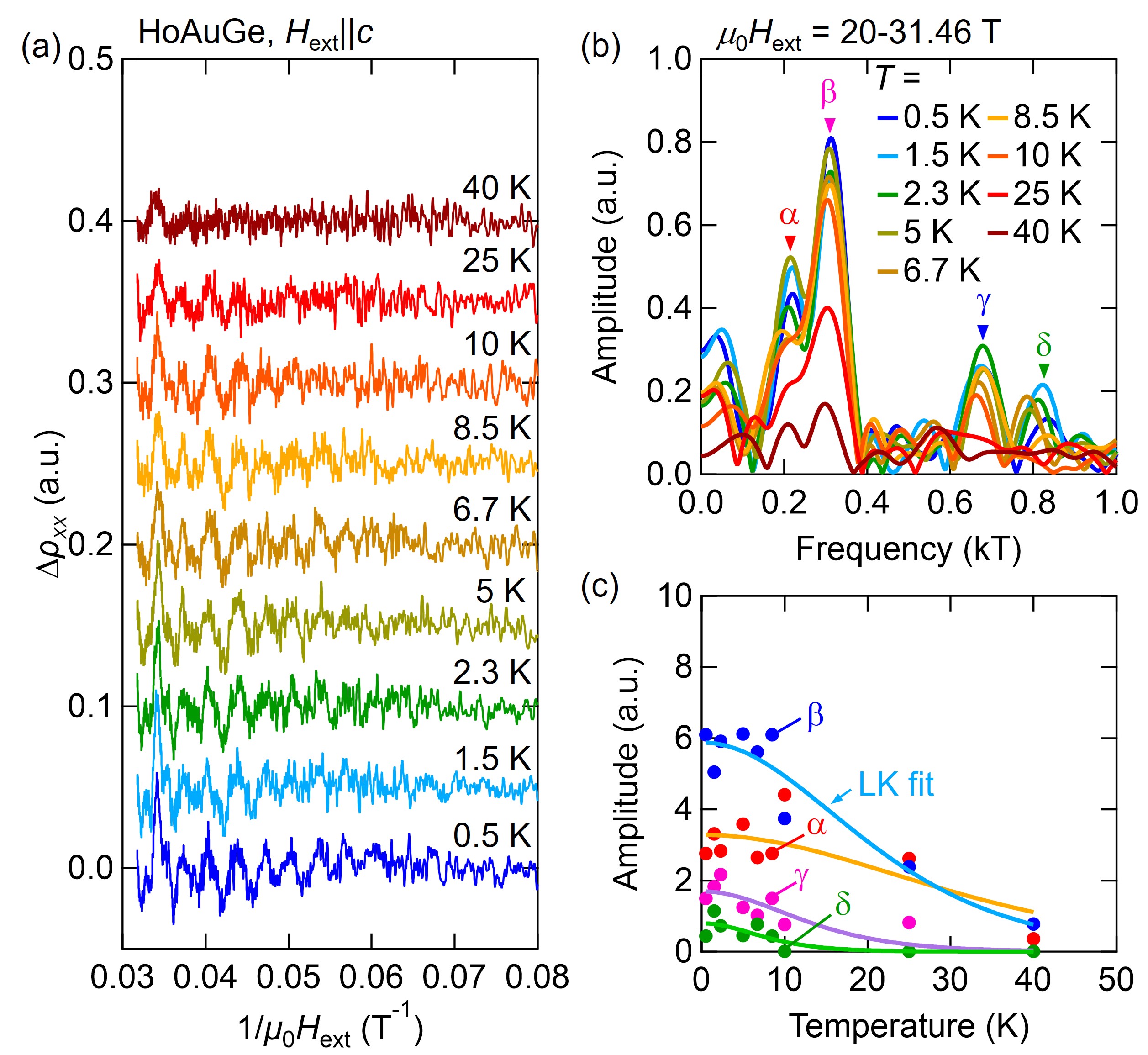}
	\caption{\label{figSTempHo}  (a) Inverse-field ($1/\mu_0H_{\text{ext}}$) dependence of the background-subtracted resistivity ($\Delta \rho_{xx}$) for HoAuGe measured at various temperatures with the $B$ along the $c$ axis.
 (b) FFT of SdH oscillations for $B\parallel c$ at various temperatures.
 Positions of the branches $\alpha$, $\beta$, $\gamma$, and $\delta$ are assigned.
 (c) Temperature dependence of peak amplitude (closed circles) and fits with LK formula (solid lines) for each branch.
}
\end{figure}

\begin{figure}[t]
	\includegraphics[width =  \columnwidth]{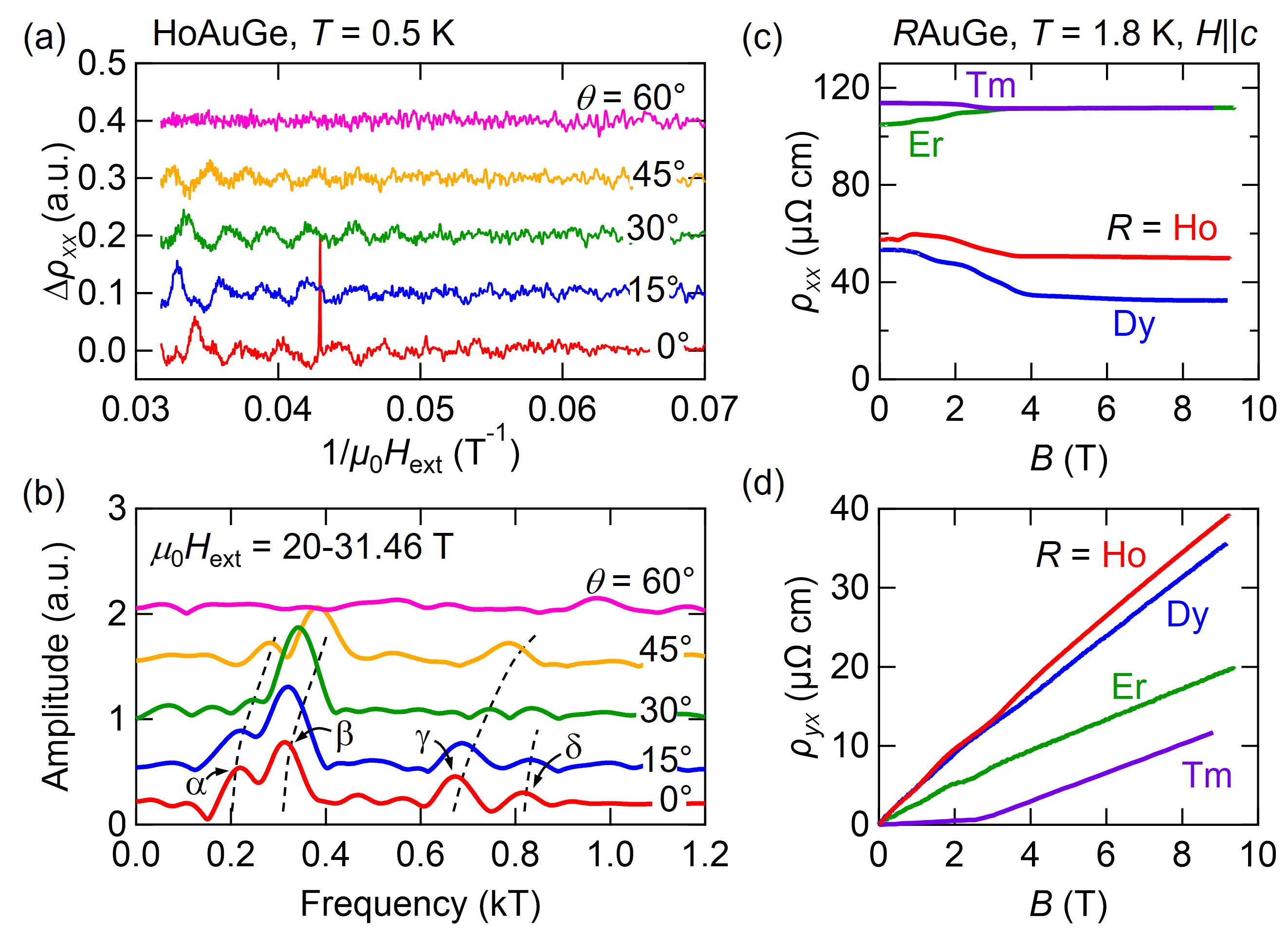}
	\caption{\label{figSAngHo} (a) Inverse-field ($1/\mu_0H_{\text{ext}}$) dependence of the background-subtracted resistivity ($\Delta \rho_{xx}$) of HoAuge, measured at $T =$ 0.5 K with the $B$ tilted from the $c$ axis to the $ab$ plane by $\theta$.
 (b) FFT of SdH oscillations at $T = 0.5$ K with various $\theta$.
 Positions of the branches $\alpha$, $\beta$, $\gamma$, and $\delta$ are assigned.
 Dashed curves are guide to eyes.
 (c)-(d) Magnetic field ($B$) dependence of (c) resistivity ($\rho_{xx}$) and (d) Hall resistivity ($\rho_{yx}$) in $R$AuGe ($R=$ Dy, Ho, Er, and Tm) for $H\parallel c$ at $T=$ 1.8 K.
 $B$ is estimated by $B=\mu_0H_{\text{ext}}-N_{\text{D}}M+M$, where $N_{\text{D}}$ is the demagnetization factor.
}
\end{figure}

Figures \ref{figSTempHo}(a)-(c) summarize the temperature dependence of SdH in HoAuGe.
Four oscillation frequencies are resolved by the FFT of $\Delta \rho_{xx}$ (Fig. \ref{figSTempHo}(b)), which are close to the frequencies corresponding to the $\alpha$, $\beta$, $\gamma$, and $\delta$ in YAuGe.
The effective mass is estimated from the temperature dependence of the oscillation amplitude (Fig. \ref{figSTempHo}(c)), which are comparable to those for YAuGe.
These results confirm the rigid band approximation with respect to the substitution of $R$ with Y.
The Fermi surface parameters are summarized in Table \ref{table}.

Figures \ref{figSAngHo}(a)-(b) show the SdH in HoAuGe at each angle of the magnetic field rotated from the $c$ axis to the $ab$ plane.
Similarly with YAuGe, the quantum oscillations shift to higher frequencies as the magnetic field is tilted towards the $ab$ plane.
Figure \ref{figBerry}(a) summarizes the angular dependence of the oscillation frequencies.

The anomalous Hall effect in DyAuGe and HoAuGe are reported in Ref. \onlinecite{kurumaji2024metamagnetism}.
We have performed the transport property measurement on ErAuGe and TmAuGe by using the single crystals obtained in our previous study~\cite{kurumaji2023single}, and have observed the anomalous magnetotransport responses.
Figures \ref{figSAngHo}(c)-(d) shows the magnetic field dependence of the resistivity in the $ab$ plane ($\rho_{xx}$) and the Hall resistivity ($\rho_{yx}$) for $R$AuGe ($R=$ Dy, Ho, Er, and Tm) in $B\parallel c$ at $T=1.8$ K.
The demagnetization effect is corrected by using the magnetization data.
Compared to $R=$ Dy and Ho, ErAuGe and TmAuGe show higher resistivity corresponding to the mobility of $\mu = 260$ cm$^2$/Vs and $\mu=160$ cm$^2$/Vs, which are estimated by the Hall coefficient ($R_{\text{H}}$) and the residual resistivity $\rho_{0}$ as $\mu=R_{\text{H}}/\rho_0$.
The Hall coefficients are estimated from the slope of $\rho_{yx}$ at the fields above the magnetization saturation, corresponding to the carrier density of $3.1\times 10^{20}$ cm$^{-3}$ for ErAuGe and $3.5\times 10^{20}$ cm$^{-3}$ for TmAuGe.

The signature of the anomalous Hall effect in ErAuGe and TmAuGe can be seen in the nonmonotonic field-dependence of $\rho_{yx}$ (Fig. \ref{figSAngHo}(d)), while the sign of the AHE in TmAuGe is negative incontrast to those in other $R$AuGe ($R=$ Dy, Ho, and Er).
The anomalous Hall conductivity ($\sigma_{xy}^{\text{A}}$) is extracted by following the procedure in Ref. \onlinecite{kurumaji2024metamagnetism}.
By using $\rho_{xx}$ and $\rho_{yx}$, the Hall conductivity is estimated as $\sigma_{xy}=\frac{\rho_{yx}}{\rho^2_{xx}+\rho^2_{yx}}$.
The anomalous Hall conductivity is obtained by the fitting the field dependence of $\sigma_{xy}$ as the summation of the normal Hall component and the AHC component proportional to the magnetization.
The field dependence of $\sigma_{xy}^{\text{A}}$ for each $R$AuGe is shown in Fig. \ref{figBerry}(f).

\begin{table*}
\caption{\label{table} Fermi surface parameters of YAuGe and HoAuGe estimated by DFT, SdH, and dHvA measurements.
$m_{\text{eff}}$: effective mass; $m_0$: free electron mass; $F$: oscillation frequency; $k_{\text{F}}$: Fermi wave number; $v_{\text{F}}$: Fermi velocity; $\tau_{\text{q}}$: quantum lifetime. $\theta$ is the angle between $B$ (or $H_{\text{ext}}$) and $c$ axis.
$k_{\text{F}}$ ($v_{\text{F}}$) is calculated by $2\pi e F/\hbar=\pi k_{\text{F}}^2$ ($v_{\text{F}}=\hbar k_{\text{F}}/m_{\text{eff}}$).
LK denotes that the parameters are obtained by the fit of the raw curves with the LK formula (Eq. (\ref{LKfull})).
Otherwise stated, the parameters are obtained by the FFT.
For $v_{\text{F}}$ of $\beta$ obtained by SdH oscillations in YAuGe, the $m_{\text{eff}}$ of $\alpha$ is used.
}
\begin{tabular}{l*{5}{c}}
\hline
\hline
Material & $\alpha$ & $\beta$ & $\gamma$ & $\delta$ \\
\hline
\hline
YAuGe (DFT, $B\parallel c$) &&&&\\
$m_{\text{eff}}/m_0$ &0.128&0.140&0.246&0.248\\
$F$ (T) &268&273&716&818\\
$k_{\text{F}}$ (\AA$^{-1}$) &0.090&0.091&0.147&0.158\\
$v_{\text{F}}$ ($10^{5}$m/s) &8.2&7.5&6.9&7.4\\
\hline

YAuGe (SdH, $B\parallel c$) &&&&\\
$m_{\text{eff}}/m_0$ &0.10& - &0.21&0.25\\
$F$ (T) &251&-&605&662\\
$F$ (T) LK &247&286&605&654\\
$k_{\text{F}}$ (\AA$^{-1}$) LK &0.087&0.093&0.136&0.141\\
$v_{\text{F}}$ ($10^{5}$m/s) LK &9.9&11&7.6&6.5\\
\hline
YAuGe (dHvA, $\theta=14^{\circ}$) &&&&\\
$m_{\text{eff}}/m_0$ & 0.13 & - &0.23&0.24\\
$F$ (T) &271&-&622&669\\
$k_{\text{F}}$ (\AA$^{-1}$) &0.091& - &0.137&0.143\\
$v_{\text{F}}$ ($10^{5}$m/s) &8.2& - &6.9&6.9\\

\hline
HoAuGe (SdH, $H_{\text{ext}}\parallel c$) &&&&\\
$m_{\text{eff}}/m_0$ & 0.12 & 0.17 & 0.28 & 0.46 \\
$F$ (T) &217&314&672&816\\
$k_{\text{F}}$ (\AA$^{-1}$) &0.081&0.098&0.143&0.157\\
$v_{\text{F}}$ ($10^{5}$m/s) &8.0&9.7&5.8&4.0\\

\hline
\hline
\end{tabular}
\end{table*}

\end{document}